\documentclass[11pt]{article}

\usepackage[preprint]{acl}

\usepackage{times}
\usepackage{latexsym}
\usepackage{comment}

\usepackage[T1]{fontenc}

\usepackage[utf8]{inputenc}
\usepackage{newunicodechar}

\newunicodechar{−}{\ensuremath{-}}

\usepackage{microtype}

\usepackage{inconsolata}

\usepackage{graphicx}
\usepackage{rotating} 
\usepackage{multirow}
\usepackage{tabularx}
\usepackage{amsmath}
\usepackage{tablefootnote}
\usepackage[table]{xcolor}
\usepackage{array}
\usepackage{hyperref}      
\usepackage{fontawesome5}  
\usepackage{xcolor}        
\renewcommand{\arraystretch}{1.4} 
\newcommand{\ha}[1]{\textcolor{red}{Hiba: #1}}

\newcommand{\xhdr}[1]{\vspace{1.7mm}\noindent{{\bf #1.}}} 

\title{Responsible Evaluation of AI for Mental Health}

\author{
\textbf{Hiba Arnaout\textsuperscript{1}}, 
\textbf{Anmol Goel\textsuperscript{1}}, 
\textbf{H. Andrew Schwartz\textsuperscript{2}}, 
\textbf{Steffen T. Eberhardt\textsuperscript{3}},\\[-1ex]
\textbf{Dana Atzil-Slonim\textsuperscript{4}}, 
\textbf{Gavin Doherty\textsuperscript{5}}, 
\textbf{Brian Schwartz\textsuperscript{3}}, 
\textbf{Wolfgang Lutz\textsuperscript{3}},\\[-1ex]
\textbf{Tim Althoff\textsuperscript{6}}, 
\textbf{Munmun De Choudhury\textsuperscript{7}}, 
\textbf{Hamidreza Jamalabadi\textsuperscript{8}}, 
\textbf{Raj Sanjay Shah\textsuperscript{7}},\\[-1ex]
\textbf{Flor Miriam Plaza-del-Arco\textsuperscript{9}}, 
\textbf{Dirk Hovy\textsuperscript{10}}, 
\textbf{Maria Liakata\textsuperscript{11}}, 
\textbf{Iryna Gurevych\textsuperscript{1}}\\
\textsuperscript{1}Technische Universität Darmstadt, 
\textsuperscript{2}Vanderbilt University
\textsuperscript{3}Trier University\\[-1ex]
\textsuperscript{4}Bar-Ilan University
\textsuperscript{5}Trinity College Dublin 
\textsuperscript{6}University of Washington\\[-1ex] 
\textsuperscript{7}Georgia Institute of Technology 
\textsuperscript{8}Phillips-Universität Marburg
\textsuperscript{9}LIACS, Leiden University\\[-1ex]
\textsuperscript{10}Bocconi University 
\textsuperscript{11}Queen Mary University London, Alan Turing Institute\\[-1ex] \\ 
}

\begin{document}
\maketitle
\begin{abstract}
Although artificial intelligence (AI) shows growing promise for mental health care, current approaches to evaluating AI tools in this domain remain fragmented and poorly aligned with clinical practice, social context, and first-hand user experience. This paper argues for a rethinking of \textit{responsible evaluation} -- what is measured, by whom, and for what purpose -- by introducing an interdisciplinary framework that integrates clinical soundness, social context, and equity, providing a structured basis for evaluation. Through an analysis of 135 recent *CL publications, we identify recurring limitations, including over-reliance on generic metrics that do not capture clinical validity, therapeutic appropriateness, or user experience, limited participation from mental health professionals, and insufficient attention to safety and equity. To address these gaps, we propose a taxonomy of AI mental health support types -- assessment-, intervention-, and information synthesis-oriented -- each with distinct risks and evaluative requirements, and illustrate its use through case studies.  
\center
{\small Project page: \url{https://ukplab.github.io/nlp-mh-evals/}}
\end{abstract}

\section{Introduction}

Large Language Models (LLMs) hold considerable promise for advancing mental health research and practice. They offer new tools at scale to support diagnosis, therapy, peer-support, and self-guided support, where users interact with LLMs directly for guidance or coping strategies~\cite{demszky2023using,cruz2025artificial}. From detecting early signs of depression in language~\cite{lan2025depression}, to clinical documentation and summarizing complex patient histories~\cite{shah-etal-2025-tn,srivastava-etal-2024-knowledge}, and generating therapeutic or supportive responses in online communities~\cite{liu2021towards,gabriel-etal-2024-ai}, AI-enabled mental health tools have the potential to augment professional care and extend psychological support beyond traditional clinical encounters. This potential is especially valuable due to the limited availability of mental health resources, growing global demand, and persistent inequities in access to care~\footnote{\href{https://www.who.int/news/item/02-09-2025-over-a-billion-people-living-with-mental-health-conditions-services-require-urgent-scale-up}{WHO 2025 report.}}.

Despite their promise, AI mental health tools are fundamentally lacking in evaluation. Existing evaluation practices are inconsistent~\cite{yang-etal-2021-weakly,aich-etal-2022-towards,chen-etal-2024-depression} and often insufficient~\cite{tornero2023methodological}. This is concerning because poor evaluation, particularly in this domain, can lead to misleading conclusions, unintended harm, and inequitable outcomes. Our review of prior AI for mental health work reveals recurring issues, including over-reliance on generic metrics that fail to capture clinical validity, therapeutic appropriateness, or user experience, minimal participation from mental health professionals, and insufficient attention to safety, equity, and long-term impact. While we do \textit{not} expect papers in venues like ACL to be fully deployable in clinical settings, careful evaluation is essential to responsibly translate research insights toward real-world mental health impact. Accordingly, our goal is to raise evaluation standards so that research outputs can earn the trust of domain experts, even when the tools are not yet -- or are not intended to be -- used in clinical practice.

These shortcomings in evaluation practices are not idiosyncratic model bugs, but symptoms of an underlying disconnect between the communities that build, use, and regulate AI for mental health tools. Current evaluations often default to technical benchmark wins, while clinicians and other users judge success by changes in symptoms, patient functioning, and safety over time; social and implementation scientists, in turn, ask whether a tool fits workflows, earns trust, and reaches people equitably. Without a shared evaluative language, results travel poorly across these communities: automated scores without clinical anchors may overstate progress, ``human studies'' may lack meaningful involvement as well as methodological transparency or expert input, and cross‑disciplinary collaboration may arrive late -- if at all. What is needed is a common, clinically grounded evaluation framework that makes psychometric constructs accessible to AI researchers, pairs them with human‑centered and implementation-science measures, and treats safety, equity, and real‑world utility as primary outcomes. This framework can then be the connective tissue that enables mutual intelligibility and, ultimately, responsible deployment across research contexts, clinics, and community platforms.

Consequently, we posit a fundamental reconsideration of evaluation for AI mental health tools according to clinical goals, typically falling into three broad types: \textbf{(1) assessment} for inferring psychological states (e.g., language-based screening), \textbf{(2) interventions} to deliver or scaffold support (e.g., therapeutic chatbots), and \textbf{(3) information synthesis} to aid practitioners or researchers (e.g., clinical summarization). Sample tasks for each type are shown in Table~\ref{tab:sampletasks}. This categorization clarifies how different types of tools require context-sensitive evaluation and enables the field to calibrate what claims are supported by existing evaluations.

\begin{table*}[ht]
\centering
\renewcommand{\arraystretch}{1.2}
\begin{tabular}{|p{3.5cm}|p{11cm}|}
\hline
\textbf{Support Type} & \textbf{Sample Tasks} \\
\hline

\textbf{Assessment} &
Depression detection from social media posts, suicide ideation risk classification, anxiety severity prediction from text or speech, emotion recognition in therapy conversations, loneliness detection.\\
\hline

\textbf{Intervention} &
Conversational CBT chatbot delivering coping strategies, sleep coaching conversational agent, mood-based coping suggestion system, guided journaling or reflection prompts, crisis de-escalation conversational support.\\
\hline

\textbf{Information synthesis} &
Therapy session summarization,  behavior coding from psychotherapy conversations, risk flagging dashboard for clinicians, symptom trend analysis and visualization, treatment recommendation support. \\
\hline

\end{tabular}
\caption{Sample tasks for each AI for mental health support type.}
\label{tab:sampletasks}
\end{table*}


\noindent
\textbf{Contributions. } Our paper makes four primary contributions. (1) We identify key gaps and challenges in current evaluation practices for AI in mental health (\S~\ref{sec:observed_practices}; see Appendix~\ref{sec:acldetails} for details of surveyed papers); (2) we propose a structured taxonomy of tool types and salient evaluation dimensions, highlighting differences between general generative AI evaluation and mental health-specific concerns (\S~\ref{sec:taxonomy}); (3) we demonstrate its utility through five illustrative case studies spanning assessment, intervention, and support tools in diverse settings (\S~\ref{sec:case_studies}); and (4), we synthesize these insights into recommendations and guiding principles for responsible and comprehensive evaluation moving forward (\S~\ref{sec:forward}).

\noindent
\textbf{Positionality.} Our call for rethinking evaluation aligns with broader reflections on the generative AI evaluation crisis in the CL community~\cite{bommasani-2023-evaluation,elangovan-etal-2024-considers,kotonya-toni-2024-towards,zhou-etal-2025-culture}, as well as work framing generative AI evaluation as a social science measurement challenge, emphasizing rigor in construct definition and validity and proposing frameworks that connect abstract evaluation goals to concrete measurement practices~\cite{wallach2025position}. While these papers focus on general-purpose generative AI, we target the clinical, ethical, and implementation challenges in mental health.

Recent surveys on LLMs in psychotherapy~\cite{na-etal-2025-survey}, cognitive distortion detection~\cite{sage-etal-2025-survey}, and mental health conversational agents~\cite{atapattu-etal-2025-exploring} primarily catalog tasks, datasets, and model capabilities, rather than providing normative guidance for responsible evaluation. Unlike~\citet{wang2025evaluating}, who review papers from 2023–2024 across medical and engineering databases to assess LLMs' clinician-like capabilities, our work surveys recent NLP research and proposes a normative, interdisciplinary evaluation framework grounded in psychometrics and clinical science.  \citet{zhang2025generative} focus on evaluating the effectiveness of generative AI chatbots through systematic review and meta-analysis, while our framework covers a broader range of AI applications in mental health, including assessment, intervention, and information synthesis. \citet{flathers2025contextualizing} propose a clinician-focused, tripartite benchmarking approach emphasizing technical safety, clinical knowledge, and reasoning, whereas our work systematically analyzes NLP research and develops a theory-grounded framework integrating psychometrics, clinical science, implementation, equity, and user experience to guide research evaluation rather than immediate clinical benchmarking.

\begin{table}[t]
\centering
\begin{tabular}{l l}
\hline
\textbf{Observed practice} & \textbf{\%}  \\
\hline
Rely \underline{only} on AI/NLP metrics& 50 \\
\underline{No} human evaluation & 52 \\
With human evaluation but \underline{no} experts & 29 \\
Evaluation guidelines \underline{not} shared & 17\\
Limitations in evaluation \underline{not} discussed & 36 \\
\hline
\end{tabular}
\caption{Overview of the ACL Anthology study conducted to ground our position. 
We queried the ACL Anthology database with mental health keywords~\tablefootnote{\textit{mental health}, \textit{mental disorder}, \textit{mental illness}, \textit{therapy} and \textit{psychiatry}; Either in the title or in the abstract.}, restricting results to the past five years and papers of types ``main'' or ``findings''. This yielded 135 papers on mental health~\tablefootnote{After manual inspection to remove papers that mention mental health \textit{only in passing} but not as the main focus; We had 152 papers before the inspection.}.
These manually-made observations provide context for our broader discussion of challenges and gaps in the evaluation of AI tools for mental health. We show details about the surveyed papers in Appendix~\ref{sec:acldetails}; Tables~\ref{tab:surveyed_works}--\ref{tab:mh_tasks}.}
\label{tab:acl_stats}
\end{table}

\section{Observed Practices}
\label{sec:observed_practices}

To ground our position, we conducted a quantitative analysis of 135  papers on mental health, published in the ACL Anthology~\footnote{\url{https://aclanthology.org/}; query date: 11-2025.} over the past 5 years, with 36\% of them published in 2025. 
The papers were coded by two annotators: one postdoc and one PhD student, both of whom have experience working on AI for mental health. 50\% of the data was double-annotated (substantial agreement; Cohen's kappa=0.67). In cases of disagreement, most of which involved inherently ambiguous instances, the senior annotator (the postdoc) conducted a more in-depth review of the paper and re-annotated the instance to reach a final decision. These ambiguities often arose in borderline cases: for example, when authors of a paper discussed limitations broadly but only briefly alluded to evaluation-related concerns without explicitly framing them as such, or when the full context needed for the annotation is buried somewhere in the Appendices. To ensure consistency in the remaining instances, the annotators discussed ambiguous cases throughout the process and made the decision together.

Table~\ref{tab:acl_stats} summarizes key patterns that emerged from this review, and Appendix~\ref{sec:acldetails} provides detailed annotations, including the tasks covered by these papers, and the observed practices documented at a paper level.   Overall, we found that current evaluation practices in this literature remain limited in scope and rigor, especially considering the sensitivity and clinical implications of the domain. While the surveyed works cover a wide range of tasks, from detecting mental health conditions~\cite{chen-etal-2024-depression,yang-etal-2021-weakly,lee-etal-2024-detecting-bipolar} to building therapeutic chatbots~\cite{saha-etal-2022-shoulder,deng-etal-2023-knowledge,shim-2021-development}, their evaluations often rely on narrow, model-centric criteria.

Specifically, five concerning patterns emerge. Half of the papers rely \textit{only} on standard AI/NLP metrics such as accuracy, F1, BLEU, or ROUGE, ignoring psychological validity or clinical relevance. Over half (54\%) include \textit{no} human evaluation, and among those that do, 29\% do so \textit{without} involving mental health experts. Nearly one-fifth of papers omit evaluation guidelines, and roughly a third fail to discuss limitations in the way the evaluations have been conducted. These gaps indicate that current practices assess technical performance but often overlook safety, interpretability, and real-world utility~\cite{thieme2020review}. 


Overall, these findings reveal a methodological gap: AI tools may score well on generic NLG metrics yet fall short of clinical standards or user needs. This critique is not aimed at individual works, but rather, highlights the need for shared, rigorous evaluation practices. The following sections build on these observations to introduce a taxonomy (\S~\ref{sec:taxonomy}), illustrate it with case studies (\S~\ref{sec:case_studies}), and present guiding principles for a clinically grounded and human-centered evaluation (\S~\ref{sec:forward}).


\begin{sidewaystable*}[htbp]
\centering
\small
\begin{tabular}{|p{3cm}|p{5cm}|p{5cm}|p{5cm}|p{5cm}|}
\hline
\cellcolor{green!20}\centering\textbf{Support type}
& \multicolumn{2}{c|}{\cellcolor{yellow!20}\textbf{Quality Criteria}}
& \multicolumn{2}{c|}{\cellcolor{blue!20}\textbf{Real-World Use}} \\
\cellcolor{green!20}
& \cellcolor{yellow!20}\centering
\textbf{Validity}\\[-2pt]
\textit{Does it do what it is intended?}
& \cellcolor{yellow!20}\centering
\textbf{Reliability}\\[-2pt]
\textit{Does it do the same thing under different conditions?}
& \cellcolor{blue!20}\centering
\textbf{Implementation}\\[-2pt]
\textit{Can it be used effectively in real-world contexts?}
& \cellcolor{blue!20}{\centering
\textbf{Maintenance}\\[-2pt]
\textit{Does it remain effective and appropriate over time as users and contexts evolve?}} \\
\hline

\cellcolor{green!20}
\textbf{Assessment} \newline
(e.g., language-based screening)
& \textbf{1. Construct Validity:} \newline
How much does it match other tools or indicators (e.g., clinical, community, or self-report measures) intended to assess the same construct (convergent) or a different construct (discriminant)? \newline
\textbf{2. Criterion Validity:} \newline
What is its association with external, theoretically-related constructs or outcomes (e.g., wellbeing, functioning, participation)?
& \textbf{1. Across Time:}  \newline What is the test-retest stability (at appropriate time intervals)? Does it [not] change if it should [not]? \newline
\textbf{2. Across Populations:}  \newline Does it work just as well across different cultures, locations, neurodivergent populations? \newline
\textbf{3. Internal Consistency:}  \newline To what extent do all components or interactions of the tool function consistently?
& \textbf{1. Feasibility:}  \newline Does it fit into the workflows and routines of intended users (e.g., clinicians, peer supporters, or individuals)? \newline
\textbf{2. Effectiveness and Usefulness} (extrinsic):  \newline Is it consistent across diverse populations? Does it improve diagnostic accuracy in practice?
\newline \textbf{3. Acceptability:}  \newline Are data gathering and feedback mechanisms for assessment acceptable to both patients and clinicians?
& \textbf{1. Generalizability and Impact:}\newline Does performance remain stable as users or contexts evolve over time? Does it contribute to improved individual or population-level outcomes? \newline
\textbf{2. Unintended Consequences:}  \newline Is it creating labeling bias? \\
\hline

\cellcolor{green!20}
\textbf{Intervention} \newline
(e.g., therapeutic chatbots)
& \textbf{1. Construct Validity:} \newline
Does it make a change in the intended direction (convergent) or have any adverse or unintended effects (discriminant)? From experimentation, RCTs, or real-world trials (efficacy or effectiveness). \newline
\textbf{2. Criterion Validity:} \newline
Does it predict or improve external downstream outcomes (e.g., wellbeing, functioning, relationships, work, community participation)?
& \textbf{1. Across Time:} \newline Does it keep working as well at future points in time? \newline
\textbf{2. Across Populations:} \newline Is the effect the same across cultures, locations, neurodivergence? \newline
\textbf{3. Internal consistency:} \newline If the intervention has multiple mechanisms or components, do they each contribute consistently to desired outcomes?
& \textbf{1. Effectiveness:} \newline Does it improve symptoms, wellbeing, or functioning under real-world conditions (with or without clinician involvement)? \newline
\textbf{2. Usability and Engagement:} \newline Do users adhere to the intervention? Is it easy to use? \newline
\textbf{3. Implementation Risk:} \newline Is it being used as intended? \newline
\textbf{4. Equity and Acceptability:} \newline Do diverse user groups find it acceptable and trustworthy? Are potential biases mitigated?
& \textbf{1. Stability:} \newline Does the benefit sustain over time across different user groups and contexts? Are there equitable outcomes and access? \newline
\textbf{2. Safety:} Are there emergent risks or harmful use patterns? \\
\hline

\cellcolor{green!20}
\textbf{Information synthesis} \newline
(e.g., clinical summarization)
& \textbf{1. Construct Validity:} \newline
Does it provide accurate, contextually appropriate, and unbiased summaries or recommendations? \newline
\textbf{2. Criterion Validity:} \newline
Does it save users (e.g., clinicians or peer supporters) time or improve the quality of their decisions?
& \textbf{1. Scenarios:} \newline Does it perform reliably across different use scenarios? \newline
\textbf{2. Services:}\newline Does it integrate effectively across different service models or modalities?
& \textbf{1. Acceptability:} \newline  Would intended users (clinicians, patients, peer supporters, community workers) accept and trust the tool in their workflows or daily lives? \newline
\textbf{2. Usefulness:} \newline Do the users find it useful in their everyday work or well-being activities? \newline
\textbf{3. Impact:} \newline Does the support improve outcomes for users or beneficiaries (e.g., efficiency, understanding, well-being)? \newline
\textbf{4. Equity and Bias Mitigation:} \newline Are there systematic biases in recommendations or summaries? Are they identified and mitigated?
& \textbf{1. Tool-level Impact:} \newline Does it reduce administrative load, emotional burden, or improve care and support quality across settings? \newline
\textbf{2. Unintended Consequences:} \newline Does it foster over-reliance or skill atrophy? \\
\hline
\end{tabular}
\caption{Taxonomy for evaluation of AI in mental health applications: aligning support types with validity, reliability, implementation, and maintenance across various contexts.}
\label{tab:taxonomy}
\end{sidewaystable*}

\section{Proposed Taxonomy}
\label{sec:taxonomy}

While new principles are needed to evaluate AI for mental health, there is much to build on from a century of work in psychological assessment (\textit{classical quantitative methods}~\cite{cook2006current}) and recent advances in applying technology in human-computer interaction and health (\textit{implementation science}~\cite{lyon2023bridging}).


\noindent
\textbf{Classical quantitative methods.} \emph{Validity} and \emph{reliability} are foundational in psychological evaluation\footnote{Evidenced by their inclusion in nearly every modern textbook on psychological research methods~\cite{cohen1988psychological,reynolds2021mastering,meyer2010understanding}}. Validity asks whether a tool \textit{does what it is intended to do}, while reliability asks \textit{whether it does so consistently}. Most current evaluations in AI mental health work mainly focuses on one validity subtype, namely construct validity, for example, through agreement with human annotations or existing scales~\cite{park2020suicidal,lee-etal-2024-detecting-bipolar}, but this is only a starting point for high-stakes applications. A classifier may correlate with overall depression severity yet fail to predict specific symptoms or generalize across populations. Similarly, a summarization tool may align with expert summaries but omit safety-critical information or misinterpret non-clinical expressions, thus highlighting limits in discriminant validity and generalization. Near-perfect construct validity is not always desirable, as even established assessments have limitations.

\noindent 
\textbf{Implementation science.} Recent advances in health informatics and human-computer interaction highlight that barriers to using AI go beyond validity and reliability~\cite{reddy2024generative}. Implementation science adds two pillars: \emph{implementation}--whether an AI tool is feasible, acceptable, fits workflows, and improves outcomes safely; and \emph{maintenance}--whether it remains effective over time, handling population shifts, language drift, inequities, or unintended consequences. Together with validity and reliability, these four pillars define a multidimensional evaluation space for AI in mental health across assessment, interventions, and information synthesis (i.e., therapist support). 


 To organize these concepts, we introduce a taxonomy of evaluation dimensions (Table~\ref{tab:taxonomy}\footnote{While our focus is on clinical integration, this taxonomy is intended to also cover peer-supported and community-based AI mental health tools.}), mapping classical psychometrics and implementation science principles onto three common AI applications: assessment, intervention, and information synthesis. These evaluation paradigms are multifaceted; no single score can capture the full opportunities and risks of AI, akin to a cockpit dashboard where multiple readings are needed to assess performance.
 
\noindent
\textbf{\emph{Assessments}} involve tools for measurement, screening, aiding diagnosis, or forecasting (e.g., scoring depression severity, estimating suicide risk from social media, classifying psychosis-related language). Validity focuses on convergent validity (alignment with other measures of the same construct), discriminant validity (avoiding spurious alignment with different constructs), and criterion validity (relation to meaningful external outcomes like hospitalization or symptom trajectories). Reliability covers stability over time (test-retest), robustness across populations (clinics, demographics, cultures, neurodivergent groups), and internal consistency (coherent subcomponents). Implementation examines feasibility, impact on diagnostic accuracy, equity, acceptability, and bias mitigation. Maintenance involves monitoring generalizability, performance drift, population-level outcomes, unintended consequences, and evolving language norms.

\noindent
\textbf{\emph{Interventions}} are tools aimed at changing outcomes, such as treatment agents, self-help aids, prevention nudges, or adaptive therapy recommendations. Validity includes construct validity (delivering the intended therapeutic ingredient), efficacy (producing beneficial change and avoiding harm), and criterion validity (predicting improvements in functioning, relationships, or job stability). Reliability examines whether effects hold across time, populations, settings, and intervention components. Implementation considers real-world symptom improvement, user engagement, clinician usability, low risk, and monitoring off-label use. Maintenance evaluates persistence of benefits and emergence of new risks, such as overuse or avoidance of human care.


\noindent
\textbf{\emph{Information synthesis}} tools augment care and administration efficiently. For automated care aids (clinical summarization, triage notes, treatment recommendations), convergent validity asks whether outputs are accurate as per the clinical evidence base, while criterion validity asks whether they save clinician time or improve documentation. Reliability emphasizes reproducibility across scenarios (note types, specialties) and modalities (telehealth vs. in-person, EHR variants). Implementation focuses on acceptability, usefulness in daily work, and patient impact.
Maintenance considers tool-level effects, like reduced burnout or unintended consequences (over-reliance, skill atrophy). 

In our evaluation framework, we prioritize these evaluation dimensions because they draw from long-standing clinical science (validity and reliability) and real-world mental health technology evaluation (implementation and maintenance), together defining the minimum requirements for responsible use in high-stakes mental health contexts.

\section{Case Studies}
\label{sec:case_studies}

The following five case studies were selected to illustrate the taxonomy across support types. They were chosen for their representativeness, methodological rigor, and the variety of AI approaches they exemplify, enabling a comprehensive demonstration of the taxonomy's dimensions: \textit{validity}, \textit{reliability}, \textit{implementation}, and \textit{maintenance}.

\subsection{Study I (\textit{Assessment}): LLM rating scales for psychometric assessment of patient engagement} 

\citet{eberhardt2025development} introduced the LLM rating scale, a psychometric tool for automatically transcribed psychotherapy sessions that measures latent psychological constructs, such as patient engagement, by applying traditional psychometric principles to AI-based assessment. The scale uses structured items--prompts like ``\textit{Please rate how motivated the patient is to engage in therapy on a scale from 0 to 100}''--to elicit zero-shot judgments from the model. The study analyzed 1,131 sessions from 155 patients using the DISCOVER framework~\cite{10.3389/fdgth.2025.1638539}, computing mean scale scores from a large pool of manually developed items, which were then evaluated for reliability and multiple forms of validity.



Validity was assessed across multiple dimensions. Construct validity was supported by moderate, significant correlations between LLM rating scale scores and engagement determinants like therapy motivation and between-session effort~\cite{HOLDSWORTH2014428}. Criterion validity was shown through associations with subsequent therapy outcomes, where higher engagement predicted greater symptom improvement. Structural validity was evaluated via multilevel confirmatory factor analysis modeling a single latent factor, with good fit ($\text{CFI} = 0.968$, $\text{SRMR} = 0.022$) though $\text{RMSEA} = 0.108$ indicated some unexplained variance. Reliability was examined as the consistency of the measurement across items, with internal consistency (McDonald's $\omega = 0.953$) showing coherent and stable LLM responses.


The study demonstrated the psychometric soundness and potential of the LLM rating scale as an automated tool for psychotherapy research and feedback. Future work should extend 
analyses across time and populations, assess robustness, fairness, and safety~\cite{lutz2024data,ryan2025fairness}, and validate across contexts, languages, and constructs. Implementation considerations, including presentation and integration (e.g., via XAI~\cite{lavelle2025explainable}), affect real-world usefulness. Testing within systems like the Trier Treatment Navigator~\cite{lutz2024data,lutz2025advances} can evaluate clinical integration and early detection potential. Ongoing maintenance is needed to monitor drift, bias, and improvements with newer LLMs.

\subsection{Study II (\textit{Assessment}): Natural language response formats for assessing depression and worry}

\citet{gu2025natural} conducted a validity- and reliability-based comparison of response formats for LLM-based assessment of depression and worry, building on prior work showing AI language assessments can approach the reliability of human scales~\cite{kjell2022natural}. The study compared four response formats, from closed to open (predefined words, descriptive words, short phrases, full-text responses), using a Sequential Evaluation with Model Pre-Registration (SEMP) design. Models were trained on a development set ($N=963$) and pre-registered before evaluation on a prospective test set ($N=145$) with validated scales, including the PHQ-9 \cite{spitzer1999validation} and GAD-7 \cite{spitzer2006brief}. The results showed strong convergent validity across formats, with correlations of $r=.60$-$.79$, exceeding the pre-registered threshold ($r>.50$). Combining all eight depression and worry models yielded correlations near or above scale reliability limits (e.g., $r=.83$ for CES-D vs.\ reliability $r=.78$), and incremental validity analyses showed improved accuracy consistent with cognitive interview theory. However, high inter-correlations among combined models ($r=.88$-$.95$) indicated reduced discriminant validity when the same responses were used to assess both constructs.

Regarding reliability and implementation, two-week test-retest correlations showed moderate to strong stability, with performance generalizing well to unseen data and prospective accuracies matching cross-validated estimates. Open-ended formats showed internal consistency at the word level, with depression- (e.g., ``blue'') and worry-related (e.g., ``anxious'') terms aligning with DSM-5 symptom clusters \cite{diagnostic2013statistical}. Implementation effectiveness was demonstrated by predicting behavioral indicators such as sick leave and mental health-related healthcare visits, often matching or exceeding standard rating scales. Feasibility analyses showed that open formats provided richer information (Shannon diversity up to 561.0) but required longer completion times (up to 4 times slower than select-word tasks).

Overall, within the proposed taxonomy, this work shows strong convergent, criterion, and external validity and temporal reliability, indicating that well-designed LLM-based assessments can rival traditional measures. However, discriminant validity and workflow feasibility remain open questions, motivating future work on cross-population reliability and integration into digital clinical platforms.

\subsection{Study III (\textit{Intervention}): Evaluating the capabilities of LLMs vs. human therapists to generate personalized interventions}

\citet{bar-shahar2025} developed an LLM-based tool for generating context-sensitive therapeutic interventions during psychotherapy sessions. It uses four specialized LLM agents, supportive, directive, exploratory, and meaning-making, along with a Judge-LLM that selects the most appropriate intervention based on the dialogue and the patient's emotional and cognitive state. This setup reflects how clinicians choose among multiple interventions and tailor responses to patients' evolving needs.

They evaluated the tool via human-AI comparisons on transcribed therapy segments, with both therapists and the AI generating interventions that expert clinicians rated for theoretical appropriateness, contextual fit, and helpfulness. High inter-rater reliability (ICC and Cohen's $\kappa$) supported robustness, and AI interventions were generally clinically relevant and sometimes approached human quality, though they lacked the depth and personalization of experienced clinicians.

Applying the taxonomy shows strong construct validity and reliability, supported by theoretical grounding and high rater agreement. However, ecological validity across therapies, languages, and contexts, fairness across groups, and key implementation and maintenance issues, such as feasibility, clinician acceptance, ethical oversight, model stability, and unintended effects, were not addressed.

Using the taxonomy, future evaluations could go beyond expert ratings by testing validity across modalities, populations, and contexts, examining reliability across models and raters, and focusing on implementation through usability, clinician-patient co-creation, and ethical integration. Ongoing maintenance would monitor drift, bias, and unintended effects, including impacts on novice therapists, ensuring the tool remains theoretically sound, reliable, and clinically sustainable.

\subsection{Study IV (\textit{Intervention}): A clinically-grounded framework for evaluating LM-assisted cognitive restructuring}

\citet{sharma-etal-2023-cognitive,sharma2024facilitating} conducted a multi-stage project to design, deploy, as well as evaluate a human-LM interaction tool for cognitive restructuring, a core Cognitive Behavioral Therapy (CBT) technique. Across all stages, the project integrated clinical validity, ecological evaluation, safety, and equity considerations.

The first stage defined and validated clinically meaningful AI objectives. Working with mental health professionals, the authors developed 7 linguistic attributes for reframing, including empathy, positivity, actionability, specificity, and addressing thinking traps. To ensure clinical validity, 600 reframes were collected and annotated by practitioners. A randomized field study ($N = 2{,}067$) on the Mental Health America platform showed users preferred highly empathic and specific reframes, while overly positive ones were less effective.

Under the implementation dimension, the framework was operationalized into an interactive LM-powered tool supporting users in cognitive restructuring. Co-designed with mental health professionals, it included safety mechanisms such as classification and rule-based filtering, IRB approval, and a user-reporting function, with flagged content (0.65\%) confirming filter effectiveness. A large-scale field study on the MHA website ($N = 15{,}531$) evaluated user-reported outcomes, including emotional impact, therapeutic utility, and skill acquisition. The tool showed measurable benefits, with the majority of participants reporting reduced negative emotion and helpfulness of reframes for overcoming negative thoughts.

Under maintenance, the third stage assessed equity and found reduced effectiveness for adolescents aged 13-17. Targeted adaptations (simpler, more casual reframes) improved helpfulness in a follow-up trial without affecting other groups, demonstrating ongoing monitoring and refinement.


This case study shows how a clinically grounded, real-world evaluation framework centered on safety and equity can produce a tool with measurable utility. From a taxonomy perspective, it shows validity (clinically aligned outcomes), reliability (consistent effects), safety (content filtering and user flagging), equity (targeted improvements), and maintenance (iterative refinement). The tool has since been deployed by Mental Health America, serving over 160,000 users~\footnote{\url{https://screening.mhanational.org/changing-thoughts-with-an-ai-assistant/}}
.

\subsection{Study V (\textit{Information synthesis}): Hierarchical LLM-VAE tool for clinically meaningful timeline summarization}

\citet{song-etal-2024-combining} proposed a hybrid tool that integrates hierarchical variational autoencoders (TH-VAEs) with LLMs to generate clinically meaningful summaries of long-term social media timelines. It produces two layers: a first-person evidence summary capturing subjective experiences, and a third-person clinical summary mapping these experiences to diagnostic indicators, interpersonal patterns, and moments of change. The goal is to help clinicians and researchers synthesize key information from longitudinal mental health data.

Evaluation of the tool integrated both automatic and expert-based components. Automatic metrics assessed meaning preservation, factual consistency, evidence appropriateness, coherence, and fluency. Clinical experts rated summaries for usefulness, diagnostic accuracy, and their ability to reflect dynamic psychological processes. Inter-rater agreement ensured the reliability of human judgments, and ablation studies tested the contribution of specific model components, such as keyphrase extraction and expert-informed prompting.

Applying the proposed taxonomy shows that the work addresses construct validity--alignment with clinical constructs--and criterion validity through correlations with expert judgments. It also touches on reliability via inter-rater agreement. However, because the tool was trained and evaluated only on social media data, ecological validity and clinical generalizability is limited. Although the peer-support platform provides authentic language, selective self-disclosure gives the model only a partial view of users' psychological states. The authors also acknowledge risks such as hallucinations, bias, and unsafe inferences, but do not systematically evaluate them, nor do they assess fairness across demographic or linguistic groups. Implementation and maintenance factors--such as clinical usability, practitioner acceptance, and long-term model stability--were likewise not examined.

Future work could extend evaluation to generalizability, usability, and sustainability. Ecological testing across cultures and clinical contexts, repeated assessments for reliability, clinician-focused implementation studies, and ongoing monitoring for drift or bias would support consistent performance, advancing the tool from proof of concept to a clinically robust, ethical, and sustainable mental health tool.

\section{Moving Forward}
\label{sec:forward}

\noindent
\textbf{Evaluation foundations and maturity pathways.} Evaluation practices in AI for mental health remain concentrated in early-stage technical validation, with relatively few tools reaching implementation or maintenance. 
While this is typical for an emerging field, it motivates the need for explicit \textit{minimum evaluation standards} appropriate for high-risk mental health contexts.
Assessment tools should demonstrate convergent and discriminant validity with clinical constructs. Intervention tools should provide evidence of therapeutic benefit, safety, and acceptability, ideally supported by prospective or randomized evaluations~\cite{hofmann2013art,cuijpers2019role}. Information synthesis tools should document measurable improvements in workflow, decision quality, or clinical comprehension.

A robust evaluation strategy requires a multilayered, standardized pipeline,  in which evaluation depth increases with a tool's intended role and potential harm. We distinguish three maturity layers:
\begin{enumerate}
    \item \textbf{Early maturity (exploratory):} Technical validation, including accuracy, robustness, and agreement with human annotations, typically using retrospective datasets. At this stage, evaluation supports feasibility assessment and hypothesis generation rather than clinical claims.
    \item \textbf{Intermediate maturity (validation):}  Human-centered evaluation, capturing expert judgment, usability, acceptability, and perceived clinical relevance, often through prospective or external validation and structured user studies.
    \item \textbf{Advanced maturity (deployment):} Assessment of contextual and ecological characteristics, including workflow integration, feasibility across settings, long-term impact, equity, safety, and monitoring of failure modes over time.
\end{enumerate}


This layered structure is particularly important in mental health settings, where concerns about invasiveness, reduced human oversight, and potential clinician deskilling are longstanding~\cite{torous2019creating}. 

Out of 60 randomly sampled papers from our set, 68\% fall into the Early Maturity (exploratory) stage, while 32\% are categorized as Intermediate Maturity (validation). Notably, more recent publications, particularly those from the past year, show a growing trend to involve clinical experts in the evaluation process.

\noindent
\textbf{Safety, fairness, and adaptability.} Safety and fairness require proactive, domain-specific protocols rather than retrospective checks. Because mental health  
involves power asymmetries and heightened risks of harm, AI support must be systematically stress-tested for hallucinations, inappropriate reassurance, and biased outputs. Fairness assessments should examine performance across demographic, cultural, and linguistic groups, acknowledging that fairness definitions entail unavoidable trade-offs~\cite{kleinberg2016inherent,ryan2025fairness}. 

Notably, most of our case studies lacked explicit safety or fairness evaluations, highlighting a significant gap for future development. Recent advances in mental health science impose additional requirements for adaptability. Clinical theory is shifting from categorical diagnoses toward dimensional and dynamic frameworks, including network-based and dynamic-systems models that conceptualize mental health states as evolving systems of interacting components~\cite{borsboom2017network,scheffer2024dynamical,ong2025llm}. AI support must therefore remain adaptable to evolving constructs and evidence, as theoretical advances directly shape evaluation targets, risk assessment, and patient safety.

\noindent
\textbf{Practical implications across maturity stages.} We emphasize that the proposed taxonomy is maturity-aware rather than a uniform checklist: early exploratory systems are not expected to satisfy deployment-level criteria such as Maintenance or full Implementation. Instead, the framework clarifies which dimensions remain unaddressed and how evaluation expectations should scale with system claims and intended use. For researchers without clinical access or deployment resources, higher-level concerns can be partially approximated through structured patient simulations, scenario-based evaluations grounded in clinical guidelines, bias audits across demographic personas, expert-informed annotation protocols targeting construct validity, and rubric-based LLM-as-a-Judge assessments aligned with clinically meaningful criteria. While such technical proxies are not substitutes for real-world validation, they enable early-stage work to engage more explicitly with safety, validity, equity, and implementation considerations, and to calibrate claims appropriately to system maturity~\cite{song2026mentalbench}.

\section{Conclusion}
This paper highlights the need to rethink how AI systems for mental health are evaluated. Our analysis of 135 *CL publications reveals recurring limitations, including reliance on generic metrics, limited involvement of mental health professionals, and insufficient attention to safety, equity, and real-world use. These gaps indicate a misalignment between current evaluation practices and the requirements of responsible deployment in mental health contexts.

To address this, we propose an interdisciplinary framework for responsible evaluation and introduce a taxonomy of AI mental health support types—assessment-, intervention-, and information synthesis-oriented—each with distinct risks and evaluation needs. This taxonomy enables evaluation practices that better reflect the intended use and risks of AI systems in mental health.


\section*{Limitations}
This paper proposes a taxonomy and accompanying evaluation framework for mental health AI, but several boundaries of scope should be noted. The analysis is informed by a set of published case studies, which may not fully represent the breadth of ongoing work or emerging AI for mental health tools. The taxonomy and evaluation pathways are conceptual rather than empirically validated, and their applicability may vary across clinical, cultural, and linguistic contexts. 

Additionally, while we outline key evaluation principles, we do not provide detailed operational metrics, leaving room for future work to refine and adapt these ideas as the field continues to develop.

\section*{Acknowledgments}
The authors acknowledge the support of Schloss
Dagstuhl – Leibniz Center for Informatics through
the Dagstuhl Seminar ‘25361: Natural Language Processing for Mental Health’.

This work was supported by the DYNAMIC Center, the LOEWE program of the Hessian Ministry of Science and Arts (Grant No. LOEWE/1/16/519/03/09.001(0009)/98), the LOEWE Distinguished Chair Ubiquitous Knowledge Processing, LOEWE initiative, Hesse, Germany (Grant No. LOEWE/4a//519/05/00.002(0002)/81), the U.S. Centers for Disease Control and Prevention/National Institute for Occupational Safety and Health (CDC/NIOSH) under Grant U01 OH012476, the German Research Foundation (Deutsche Forschungsgemeinschaft, DFG; Grant Nos. 493169211 and 525286173), the National Institutes of Health through Grants P50MH115838, R01MH117172, and R01MH135488, the American Foundation for Suicide Prevention and the Betty and Gordon Moore Foundation, the Excellence Cluster EXC3066 The Adaptive Mind, the European Research Council (ERC) under the European Union's Horizon 2020 research and innovation programme (Grant Agreement No. 949944, INTEGRATOR), the MUR FARE 2020 initiative (Grant Agreement Prot. R20YSMBZ8S, INDOMITA), the Data and Marketing Insights Unit of the Bocconi Institute for Data Science and Analysis (BIDSA), the Keystone grant funding from Responsible AI UK (Grant No. EP/Y009800/1), and the Research Ireland Adapt Research Centre (grant 13/RC/2106\_P2).

\bibliography{custom}

@inproceedings{sharma2024facilitating,
  title={Facilitating self-guided mental health interventions through human-language model interaction: A case study of cognitive restructuring},
  author={Sharma, Ashish and Rushton, Kevin and Lin, Inna Wanyin and Nguyen, Theresa and Althoff, Tim},
  booktitle={Proceedings of the 2024 CHI Conference on Human Factors in Computing Systems},
  pages={1--29},
  year={2024},
url={https://dl.acm.org/doi/full/10.1145/3613904.3642761}
}

@article{thieme2020review,
author = {Thieme, Anja and Belgrave, Danielle and Doherty, Gavin},
title = {Machine Learning in Mental Health: A Systematic Review of the {HCI} Literature to Support the Development of Effective and Implementable {ML} Systems},
year = {2020},
issue_date = {October 2020},
publisher = {Association for Computing Machinery},
address = {New York, NY, USA},
volume = {27},
number = {5},
issn = {1073-0516},
url = {https://doi.org/10.1145/3398069},
doi = {10.1145/3398069},
journal = {ACM Transactions on Computer-Human Interaction},
month = aug,
articleno = {34},
numpages = {53}
}

@article{ryan2025fairness,
author = {Ryan, Seamus and Cai, Wanling and Bowman, Robert and Doherty, Gavin},
title = {Fairness Challenges in the Design of Machine Learning Applications for Healthcare},
year = {2025},
issue_date = {October 2025},
publisher = {Association for Computing Machinery},
address = {New York, NY, USA},
volume = {6},
number = {4},
url = {https://doi.org/10.1145/3728368},
doi = {10.1145/3728368},
journal = {ACM Trans. Comput. Healthcare},
month = oct,
articleno = {40},
numpages = {26}
}

@inproceedings{bommasani-2023-evaluation,
    title = "Evaluation for Change",
    author = "Bommasani, Rishi",
    editor = "Rogers, Anna  and
      Boyd-Graber, Jordan  and
      Okazaki, Naoaki",
    booktitle = "Findings of the Association for Computational Linguistics: ACL 2023",
    month = jul,
    year = "2023",
    address = "Toronto, Canada",
    publisher = "Association for Computational Linguistics",
    url = "https://aclanthology.org/2023.findings-acl.522/",
    doi = "10.18653/v1/2023.findings-acl.522",
    pages = "8227--8239"
}

@inproceedings{elangovan-etal-2024-considers,
    title = "{C}on{S}i{DERS}-The-Human Evaluation Framework: Rethinking Human Evaluation for Generative Large Language Models",
    author = "Elangovan, Aparna  and
      Liu, Ling  and
      Xu, Lei  and
      Bodapati, Sravan Babu  and
      Roth, Dan",
    editor = "Ku, Lun-Wei  and
      Martins, Andre  and
      Srikumar, Vivek",
    booktitle = "Proceedings of the 62nd Annual Meeting of the Association for Computational Linguistics (Volume 1: Long Papers)",
    month = aug,
    year = "2024",
    address = "Bangkok, Thailand",
    publisher = "Association for Computational Linguistics",
    url = "https://aclanthology.org/2024.acl-long.63/",
    doi = "10.18653/v1/2024.acl-long.63",
    pages = "1137--1160"
}

@inproceedings{nguyen-etal-2024-taking,
    title = "Taking a turn for the better: Conversation redirection throughout the course of mental-health therapy",
    author = "Nguyen, Vivian  and
      Jung, Sang Min  and
      Lee, Lillian  and
      Hull, Thomas D.  and
      Danescu-Niculescu-Mizil, Cristian",
    editor = "Al-Onaizan, Yaser  and
      Bansal, Mohit  and
      Chen, Yun-Nung",
    booktitle = "Findings of the Association for Computational Linguistics: EMNLP 2024",
    month = nov,
    year = "2024",
    address = "Miami, Florida, USA",
    publisher = "Association for Computational Linguistics",
    url = "https://aclanthology.org/2024.findings-emnlp.555/",
    doi = "10.18653/v1/2024.findings-emnlp.555",
    pages = "9507--9521"
}

@article{lutz2025advances,
  title={Advances in personalization of psychological interventions},
  author={Lutz, Wolfgang and Schwartz, Brian and Vehlen, Antonia and Eberhardt, Steffen T and Delgadillo, Jaime},
  journal={World Psychiatry},
  volume={24},
  number={3},
  pages={343},
  year={2025},
url={https://pmc.ncbi.nlm.nih.gov/articles/PMC12434349/}
}

@article{gu2025natural,
   title={Natural language response formats for assessing depression and worry with large language models: A sequential evaluation with model pre-registration},
   author={Gu, Zhuojun and Kjell, Katarina and Schwartz, H Andrew and Kjell, Oscar},
   journal={Assessment},
   pages={10731911251364022},
   year={2025},
   publisher={SAGE Publications Sage CA: Los Angeles, CA},
url={https://journals.sagepub.com/doi/full/10.1177/10731911251364022}
 }

@article{lutz2024data,
  title={Data-informed psychological therapy, measurement-based care, and precision mental health.},
  author={Lutz, Wolfgang and Vehlen, Antonia and Schwartz, Brian},
  journal={Journal of Consulting and Clinical Psychology},
  volume={92},
  number={10},
  pages={671},
  year={2024},
  publisher={American Psychological Association},
url={https://psycnet.apa.org/record/2025-41436-001}
}

@article{HOLDSWORTH2014428,
title = {Client engagement in psychotherapeutic treatment and associations with client characteristics, therapist characteristics, and treatment factors},
journal = {Clinical Psychology Review},
volume = {34},
number = {5},
pages = {428-450},
year = {2014},
issn = {0272-7358},
doi = {https://doi.org/10.1016/j.cpr.2014.06.004},
url = {https://www.sciencedirect.com/science/article/pii/S0272735814000968},
author = {Emma Holdsworth and Erica Bowen and Sarah Brown and Douglas Howat}
}

@ARTICLE{10.3389/fdgth.2025.1638539,
  
AUTHOR={Hallmen, Tobias  and Schiller, Dominik  and Vehlen, Antonia  and Eberhardt, Steffen  and Baur, Tobias  and Withanage Don, Daksitha  and Lutz, Wolfgang  and André, Elisabeth },
         
TITLE={DISCOVER: a Data-driven Interactive System for Comprehensive Observation, Visualization, and ExploRation of human behavior},
        
JOURNAL={Frontiers in Digital Health},
        
VOLUME={Volume 7 - 2025},

YEAR={2025},

URL={https://www.frontiersin.org/journals/digital-health/articles/10.3389/fdgth.2025.1638539},

DOI={10.3389/fdgth.2025.1638539},

ISSN={2673-253X}}

@article{eberhardt2025development,
  title={Development and validation of large language model rating scales for automatically transcribed psychological therapy sessions},
  author={Eberhardt, Steffen T and Vehlen, Antonia and Schaffrath, Jana and Schwartz, Brian and Baur, Tobias and Schiller, Dominik and Hallmen, Tobias and Andr{\'e}, Elisabeth and Lutz, Wolfgang},
  journal={Scientific Reports},
  volume={15},
  number={1},
  pages={29541},
  year={2025},
  publisher={Nature Publishing Group UK London},
url={https://www.nature.com/articles/s41598-025-14923-y}
}

@article{kjell2022natural,
  title={Natural language analyzed with AI-based transformers predict traditional subjective well-being measures approaching the theoretical upper limits in accuracy},
  author={Kjell, Oscar NE and Sikstr{\"o}m, Sverker and Kjell, Katarina and Schwartz, H Andrew},
  journal={Scientific reports},
  volume={12},
  number={1},
  pages={3918},
  year={2022},
  publisher={Nature Publishing Group UK London},
url={https://www.nature.com/articles/s41598-022-07520-w}
}

@inproceedings{kotonya-toni-2024-towards,
    title = "Towards a Framework for Evaluating Explanations in Automated Fact Verification",
    author = "Kotonya, Neema  and
      Toni, Francesca",
    editor = "Calzolari, Nicoletta  and
      Kan, Min-Yen  and
      Hoste, Veronique  and
      Lenci, Alessandro  and
      Sakti, Sakriani  and
      Xue, Nianwen",
    booktitle = "Proceedings of the 2024 Joint International Conference on Computational Linguistics, Language Resources and Evaluation (LREC-COLING 2024)",
    month = may,
    year = "2024",
    address = "Torino, Italia",
    publisher = "ELRA and ICCL",
    url = "https://aclanthology.org/2024.lrec-main.1422/",
    pages = "16364--16377"
}

@inproceedings{garg-etal-2023-annotated,
    title = "An Annotated Dataset for Explainable Interpersonal Risk Factors of Mental Disturbance in Social Media Posts",
    author = "Garg, Muskan  and
      Shahbandegan, Amirmohammad  and
      Chadha, Amrit  and
      Mago, Vijay",
    editor = "Rogers, Anna  and
      Boyd-Graber, Jordan  and
      Okazaki, Naoaki",
    booktitle = "Findings of the Association for Computational Linguistics: ACL 2023",
    month = jul,
    year = "2023",
    address = "Toronto, Canada",
    publisher = "Association for Computational Linguistics",
    url = "https://aclanthology.org/2023.findings-acl.757/",
    doi = "10.18653/v1/2023.findings-acl.757",
    pages = "11960--11969"
}

@inproceedings{chandra-etal-2025-lived,
    title = "Lived Experience Not Found: {LLM}s Struggle to Align with Experts on Addressing Adverse Drug Reactions from Psychiatric Medication Use",
    author = "Chandra, Mohit  and
      Sriraman, Siddharth  and
      Verma, Gaurav  and
      Khanuja, Harneet Singh  and
      Campayo, Jose Suarez  and
      Li, Zihang  and
      Birnbaum, Michael L.  and
      De Choudhury, Munmun",
    editor = "Chiruzzo, Luis  and
      Ritter, Alan  and
      Wang, Lu",
    booktitle = "Proceedings of the 2025 Conference of the Nations of the Americas Chapter of the Association for Computational Linguistics: Human Language Technologies (Volume 1: Long Papers)",
    month = apr,
    year = "2025",
    address = "Albuquerque, New Mexico",
    publisher = "Association for Computational Linguistics",
    url = "https://aclanthology.org/2025.naacl-long.553/",
    doi = "10.18653/v1/2025.naacl-long.553",
    pages = "11083--11113",
    ISBN = "979-8-89176-189-6"
}

@inproceedings{yang-etal-2021-weakly,
    title = "Weakly-Supervised Methods for Suicide Risk Assessment: Role of Related Domains",
    author = "Yang, Chenghao  and
      Zhang, Yudong  and
      Muresan, Smaranda",
    editor = "Zong, Chengqing  and
      Xia, Fei  and
      Li, Wenjie  and
      Navigli, Roberto",
    booktitle = "Proceedings of the 59th Annual Meeting of the Association for Computational Linguistics and the 11th International Joint Conference on Natural Language Processing (Volume 2: Short Papers)",
    month = aug,
    year = "2021",
    address = "Online",
    publisher = "Association for Computational Linguistics",
    url = "https://aclanthology.org/2021.acl-short.133/",
    doi = "10.18653/v1/2021.acl-short.133",
    pages = "1049--1057"
}

@inproceedings{bouzoubaa-etal-2024-decoding,
    title = "Decoding the Narratives: Analyzing Personal Drug Experiences Shared on {R}eddit",
    author = "Bouzoubaa, Layla  and
      Aghakhani, Elham  and
      Song, Max  and
      Trinh, Quang  and
      Rezapour, Shadi",
    editor = "Ku, Lun-Wei  and
      Martins, Andre  and
      Srikumar, Vivek",
    booktitle = "Findings of the Association for Computational Linguistics: ACL 2024",
    month = aug,
    year = "2024",
    address = "Bangkok, Thailand",
    publisher = "Association for Computational Linguistics",
    url = "https://aclanthology.org/2024.findings-acl.367/",
    doi = "10.18653/v1/2024.findings-acl.367",
    pages = "6131--6148"
}

@inproceedings{lissak-etal-2024-colorful,
    title = "The Colorful Future of {LLM}s: Evaluating and Improving {LLM}s as Emotional Supporters for Queer Youth",
    author = "Lissak, Shir  and
      Calderon, Nitay  and
      Shenkman, Geva  and
      Ophir, Yaakov  and
      Fruchter, Eyal  and
      Brunstein Klomek, Anat  and
      Reichart, Roi",
    editor = "Duh, Kevin  and
      Gomez, Helena  and
      Bethard, Steven",
    booktitle = "Proceedings of the 2024 Conference of the North American Chapter of the Association for Computational Linguistics: Human Language Technologies (Volume 1: Long Papers)",
    month = jun,
    year = "2024",
    address = "Mexico City, Mexico",
    publisher = "Association for Computational Linguistics",
    url = "https://aclanthology.org/2024.naacl-long.113/",
    doi = "10.18653/v1/2024.naacl-long.113",
    pages = "2040--2079"
}

@inproceedings{zhang-etal-2024-llms,
    title = "When {LLM}s Meets Acoustic Landmarks: An Efficient Approach to Integrate Speech into Large Language Models for Depression Detection",
    author = "Zhang, Xiangyu  and
      Liu, Hexin  and
      Xu, Kaishuai  and
      Zhang, Qiquan  and
      Liu, Daijiao  and
      Ahmed, Beena  and
      Epps, Julien",
    editor = "Al-Onaizan, Yaser  and
      Bansal, Mohit  and
      Chen, Yun-Nung",
    booktitle = "Proceedings of the 2024 Conference on Empirical Methods in Natural Language Processing",
    month = nov,
    year = "2024",
    address = "Miami, Florida, USA",
    publisher = "Association for Computational Linguistics",
    url = "https://aclanthology.org/2024.emnlp-main.8/",
    doi = "10.18653/v1/2024.emnlp-main.8",
    pages = "146--158"
}

@inproceedings{chen-etal-2023-detection,
    title = "Detection of Multiple Mental Disorders from Social Media with Two-Stream Psychiatric Experts",
    author = "Chen, Siyuan  and
      Zhang, Zhiling  and
      Wu, Mengyue  and
      Zhu, Kenny",
    editor = "Bouamor, Houda  and
      Pino, Juan  and
      Bali, Kalika",
    booktitle = "Proceedings of the 2023 Conference on Empirical Methods in Natural Language Processing",
    month = dec,
    year = "2023",
    address = "Singapore",
    publisher = "Association for Computational Linguistics",
    url = "https://aclanthology.org/2023.emnlp-main.562/",
    doi = "10.18653/v1/2023.emnlp-main.562",
    pages = "9071--9084"
}

@inproceedings{gogoulou-etal-2021-predicting,
    title = "Predicting Treatment Outcome from Patient Texts:The Case of {I}nternet-Based Cognitive Behavioural Therapy",
    author = "Gogoulou, Evangelia  and
      Boman, Magnus  and
      Ben Abdesslem, Fehmi  and
      Hentati Isacsson, Nils  and
      Kaldo, Viktor  and
      Sahlgren, Magnus",
    editor = "Merlo, Paola  and
      Tiedemann, Jorg  and
      Tsarfaty, Reut",
    booktitle = "Proceedings of the 16th Conference of the European Chapter of the Association for Computational Linguistics: Main Volume",
    month = apr,
    year = "2021",
    address = "Online",
    publisher = "Association for Computational Linguistics",
    url = "https://aclanthology.org/2021.eacl-main.46/",
    doi = "10.18653/v1/2021.eacl-main.46",
    pages = "575--580"
}

@inproceedings{mishra-etal-2023-pal,
    title = "{PAL} to Lend a Helping Hand: Towards Building an Emotion Adaptive Polite and Empathetic Counseling Conversational Agent",
    author = "Mishra, Kshitij  and
      Priya, Priyanshu  and
      Ekbal, Asif",
    editor = "Rogers, Anna  and
      Boyd-Graber, Jordan  and
      Okazaki, Naoaki",
    booktitle = "Proceedings of the 61st Annual Meeting of the Association for Computational Linguistics (Volume 1: Long Papers)",
    month = jul,
    year = "2023",
    address = "Toronto, Canada",
    publisher = "Association for Computational Linguistics",
    url = "https://aclanthology.org/2023.acl-long.685/",
    doi = "10.18653/v1/2023.acl-long.685",
    pages = "12254--12271"
}

@inproceedings{xiao-etal-2024-healme,
    title = "{H}eal{M}e: Harnessing Cognitive Reframing in Large Language Models for Psychotherapy",
    author = "Xiao, Mengxi  and
      Xie, Qianqian  and
      Kuang, Ziyan  and
      Liu, Zhicheng  and
      Yang, Kailai  and
      Peng, Min  and
      Han, Weiguang  and
      Huang, Jimin",
    editor = "Ku, Lun-Wei  and
      Martins, Andre  and
      Srikumar, Vivek",
    booktitle = "Proceedings of the 62nd Annual Meeting of the Association for Computational Linguistics (Volume 1: Long Papers)",
    month = aug,
    year = "2024",
    address = "Bangkok, Thailand",
    publisher = "Association for Computational Linguistics",
    url = "https://aclanthology.org/2024.acl-long.93/",
    doi = "10.18653/v1/2024.acl-long.93",
    pages = "1707--1725"
}

@inproceedings{wang-etal-2023-c2d2,
    title = "{C}2{D}2 Dataset: A Resource for the Cognitive Distortion Analysis and Its Impact on Mental Health",
    author = "Wang, Bichen  and
      Deng, Pengfei  and
      Zhao, Yanyan  and
      Qin, Bing",
    editor = "Bouamor, Houda  and
      Pino, Juan  and
      Bali, Kalika",
    booktitle = "Findings of the Association for Computational Linguistics: EMNLP 2023",
    month = dec,
    year = "2023",
    address = "Singapore",
    publisher = "Association for Computational Linguistics",
    url = "https://aclanthology.org/2023.findings-emnlp.680/",
    doi = "10.18653/v1/2023.findings-emnlp.680",
    pages = "10149--10160"
}

@inproceedings{lozoya-etal-2024-generating,
    title = "Generating Mental Health Transcripts with {SAPE} ({S}panish Adaptive Prompt Engineering)",
    author = "Lozoya, Daniel  and
      Berazaluce, Alejandro  and
      Perches, Juan  and
      L{\'u}a, Eloy  and
      Conway, Mike  and
      D{'}Alfonso, Simon",
    editor = "Duh, Kevin  and
      Gomez, Helena  and
      Bethard, Steven",
    booktitle = "Proceedings of the 2024 Conference of the North American Chapter of the Association for Computational Linguistics: Human Language Technologies (Volume 1: Long Papers)",
    month = jun,
    year = "2024",
    address = "Mexico City, Mexico",
    publisher = "Association for Computational Linguistics",
    url = "https://aclanthology.org/2024.naacl-long.285/",
    doi = "10.18653/v1/2024.naacl-long.285",
    pages = "5096--5113"
}

@article{wallach2025position,
  title={Position: Evaluating generative ai systems is a social science measurement challenge},
  author={Wallach, Hanna and Desai, Meera and Cooper, A Feder and Wang, Angelina and Atalla, Chad and Barocas, Solon and Blodgett, Su Lin and Chouldechova, Alexandra and Corvi, Emily and Dow, P Alex and others},
  journal={arXiv preprint arXiv:2502.00561},
  year={2025}
}

@article{flathers2025contextualizing,
  title={Contextualizing clinical benchmarks: a tripartite approach to evaluating LLM-based tools in mental health settings},
  author={Flathers, Matthew and Dwyer, Bridget and Rozenblit, Eden and Torous, John},
  journal={Journal of Psychiatric Practice{\textregistered}},
  volume={31},
  number={6},
  pages={294--301},
  year={2025},
  publisher={LWW}
}

@article{zhang2025generative,
  title={Generative AI Mental Health Chatbots as Therapeutic Tools: Systematic Review and Meta-Analysis of Their Role in Reducing Mental Health Issues},
  author={Zhang, Qiyang and Zhang, Renwen and Xiong, Yiying and Sui, Yuan and Tong, Chang and Lin, Fu-Hung},
  journal={Journal of Medical Internet Research},
  volume={27},
  pages={e78238},
  year={2025},
  publisher={JMIR Publications Toronto, Canada}
}

@article{wang2025evaluating,
  title={Evaluating generative AI in mental health: systematic review of capabilities and limitations},
  author={Wang, Liying and Bhanushali, Tanmay and Huang, Zhuoran and Yang, Jingyi and Badami, Sukriti and Hightow-Weidman, Lisa},
  journal={JMIR mental health},
  volume={12},
  number={1},
  pages={e70014},
  year={2025},
  publisher={JMIR Publications Inc., Toronto, Canada}
}

@inproceedings{wei2021linguistic,
    title = "Linguistic Complexity Loss in Text-Based Therapy",
    author = "Wei, Jason  and
      Finn, Kelly  and
      Templeton, Emma  and
      Wheatley, Thalia  and
      Vosoughi, Soroush",
    editor = "Toutanova, Kristina  and
      Rumshisky, Anna  and
      Zettlemoyer, Luke  and
      Hakkani-Tur, Dilek  and
      Beltagy, Iz  and
      Bethard, Steven  and
      Cotterell, Ryan  and
      Chakraborty, Tanmoy  and
      Zhou, Yichao",
    booktitle = "Proceedings of the 2021 Conference of the North American Chapter of the Association for Computational Linguistics: Human Language Technologies",
    month = jun,
    year = "2021",
    address = "Online",
    publisher = "Association for Computational Linguistics",
    url = "https://aclanthology.org/2021.naacl-main.352/",
    doi = "10.18653/v1/2021.naacl-main.352",
    pages = "4450--4459"
}

@inproceedings{zhou-etal-2025-culture,
    title = "Culture is Not Trivia: Sociocultural Theory for Cultural {NLP}",
    author = "Zhou, Naitian  and
      Bamman, David  and
      Bleaman, Isaac L.",
    editor = "Che, Wanxiang  and
      Nabende, Joyce  and
      Shutova, Ekaterina  and
      Pilehvar, Mohammad Taher",
    booktitle = "Proceedings of the 63rd Annual Meeting of the Association for Computational Linguistics (Volume 1: Long Papers)",
    month = jul,
    year = "2025",
    address = "Vienna, Austria",
    publisher = "Association for Computational Linguistics",
    url = "https://aclanthology.org/2025.acl-long.1256/",
    doi = "10.18653/v1/2025.acl-long.1256",
    pages = "25869--25886",
    ISBN = "979-8-89176-251-0"
}

@article{demszky2023using,
  title={Using large language models in psychology},
  author={Demszky, Dorottya and Yang, Diyi and Yeager, David S and Bryan, Christopher J and Clapper, Margarett and Chandhok, Susannah and Eichstaedt, Johannes C and Hecht, Cameron and Jamieson, Jeremy and Johnson, Meghann and others},
  journal={Nature Reviews Psychology},
  volume={2},
  number={11},
  pages={688--701},
  year={2023},
  publisher={Nature Publishing Group US New York}
}

@inproceedings{chen-etal-2023-empowering,
    title = "Empowering Psychotherapy with Large Language Models: Cognitive Distortion Detection through Diagnosis of Thought Prompting",
    author = "Chen, Zhiyu  and
      Lu, Yujie  and
      Wang, William",
    editor = "Bouamor, Houda  and
      Pino, Juan  and
      Bali, Kalika",
    booktitle = "Findings of the Association for Computational Linguistics: EMNLP 2023",
    month = dec,
    year = "2023",
    address = "Singapore",
    publisher = "Association for Computational Linguistics",
    url = "https://aclanthology.org/2023.findings-emnlp.284/",
    doi = "10.18653/v1/2023.findings-emnlp.284",
    pages = "4295--4304"
}

@inproceedings{zhang-etal-2023-ask,
    title = "Ask an Expert: Leveraging Language Models to Improve Strategic Reasoning in Goal-Oriented Dialogue Models",
    author = "Zhang, Qiang  and
      Naradowsky, Jason  and
      Miyao, Yusuke",
    editor = "Rogers, Anna  and
      Boyd-Graber, Jordan  and
      Okazaki, Naoaki",
    booktitle = "Findings of the Association for Computational Linguistics: ACL 2023",
    month = jul,
    year = "2023",
    address = "Toronto, Canada",
    publisher = "Association for Computational Linguistics",
    url = "https://aclanthology.org/2023.findings-acl.417/",
    doi = "10.18653/v1/2023.findings-acl.417",
    pages = "6665--6694"
}

@inproceedings{zanwar-etal-2023-fuse,
    title = "What to Fuse and How to Fuse: Exploring Emotion and Personality Fusion Strategies for Explainable Mental Disorder Detection",
    author = "Zanwar, Sourabh  and
      Li, Xiaofei  and
      Wiechmann, Daniel  and
      Qiao, Yu  and
      Kerz, Elma",
    editor = "Rogers, Anna  and
      Boyd-Graber, Jordan  and
      Okazaki, Naoaki",
    booktitle = "Findings of the Association for Computational Linguistics: ACL 2023",
    month = jul,
    year = "2023",
    address = "Toronto, Canada",
    publisher = "Association for Computational Linguistics",
    url = "https://aclanthology.org/2023.findings-acl.568/",
    doi = "10.18653/v1/2023.findings-acl.568",
    pages = "8926--8940"
}

@inproceedings{nguyen-etal-2022-improving,
    title = "Improving the Generalizability of Depression Detection by Leveraging Clinical Questionnaires",
    author = "Nguyen, Thong  and
      Yates, Andrew  and
      Zirikly, Ayah  and
      Desmet, Bart  and
      Cohan, Arman",
    editor = "Muresan, Smaranda  and
      Nakov, Preslav  and
      Villavicencio, Aline",
    booktitle = "Proceedings of the 60th Annual Meeting of the Association for Computational Linguistics (Volume 1: Long Papers)",
    month = may,
    year = "2022",
    address = "Dublin, Ireland",
    publisher = "Association for Computational Linguistics",
    url = "https://aclanthology.org/2022.acl-long.578/",
    doi = "10.18653/v1/2022.acl-long.578",
    pages = "8446--8459"
}

@inproceedings{sawhney-etal-2021-phase,
    title = "{PHASE}: Learning Emotional Phase-aware Representations for Suicide Ideation Detection on Social Media",
    author = "Sawhney, Ramit  and
      Joshi, Harshit  and
      Flek, Lucie  and
      Shah, Rajiv Ratn",
    editor = "Merlo, Paola  and
      Tiedemann, Jorg  and
      Tsarfaty, Reut",
    booktitle = "Proceedings of the 16th Conference of the European Chapter of the Association for Computational Linguistics: Main Volume",
    month = apr,
    year = "2021",
    address = "Online",
    publisher = "Association for Computational Linguistics",
    url = "https://aclanthology.org/2021.eacl-main.205/",
    doi = "10.18653/v1/2021.eacl-main.205",
    pages = "2415--2428"
}

@inproceedings{abdelkadir-etal-2024-diverse,
    title = "Diverse Perspectives, Divergent Models: Cross-Cultural Evaluation of Depression Detection on {T}witter",
    author = "Abdelkadir, Nuredin Ali  and
      Zhang, Charles  and
      Mayo, Ned  and
      Chancellor, Stevie",
    editor = "Duh, Kevin  and
      Gomez, Helena  and
      Bethard, Steven",
    booktitle = "Proceedings of the 2024 Conference of the North American Chapter of the Association for Computational Linguistics: Human Language Technologies (Volume 2: Short Papers)",
    month = jun,
    year = "2024",
    address = "Mexico City, Mexico",
    publisher = "Association for Computational Linguistics",
    url = "https://aclanthology.org/2024.naacl-short.58/",
    doi = "10.18653/v1/2024.naacl-short.58",
    pages = "672--680"
}

@inproceedings{li-etal-2024-understanding-therapeutic,
    title = "Understanding the Therapeutic Relationship between Counselors and Clients in Online Text-based Counseling using {LLM}s",
    author = "Li, Anqi  and
      Lu, Yu  and
      Song, Nirui  and
      Zhang, Shuai  and
      Ma, Lizhi  and
      Lan, Zhenzhong",
    editor = "Al-Onaizan, Yaser  and
      Bansal, Mohit  and
      Chen, Yun-Nung",
    booktitle = "Findings of the Association for Computational Linguistics: EMNLP 2024",
    month = nov,
    year = "2024",
    address = "Miami, Florida, USA",
    publisher = "Association for Computational Linguistics",
    url = "https://aclanthology.org/2024.findings-emnlp.69/",
    doi = "10.18653/v1/2024.findings-emnlp.69",
    pages = "1280--1303"
}

@inproceedings{gollapalli-etal-2023-identifying,
    title = "Identifying {Early Maladaptive Schemas} from Mental Health Question Texts",
    author = "Gollapalli, Sujatha  and
      Ang, Beng  and
      Ng, See-Kiong",
    editor = "Bouamor, Houda  and
      Pino, Juan  and
      Bali, Kalika",
    booktitle = "Findings of the Association for Computational Linguistics: EMNLP 2023",
    month = dec,
    year = "2023",
    address = "Singapore",
    publisher = "Association for Computational Linguistics",
    url = "https://aclanthology.org/2023.findings-emnlp.792/",
    doi = "10.18653/v1/2023.findings-emnlp.792",
    pages = "11832--11843"
}

@inproceedings{kian-etal-2025-using,
    title = "Using Linguistic Entrainment to Evaluate Large Language Models for Use in Cognitive Behavioral Therapy",
    author = "Kian, Mina  and
      Shrestha, Kaleen  and
      Fischer, Katrin  and
      Zhu, Xiaoyuan  and
      Ong, Jonathan  and
      Trehan, Aryan  and
      Wang, Jessica  and
      Chang, Gloria  and
      Arnold, S{\'e}b  and
      Mataric, Maja",
    editor = "Chiruzzo, Luis  and
      Ritter, Alan  and
      Wang, Lu",
    booktitle = "Findings of the Association for Computational Linguistics: NAACL 2025",
    month = apr,
    year = "2025",
    address = "Albuquerque, New Mexico",
    publisher = "Association for Computational Linguistics",
    url = "https://aclanthology.org/2025.findings-naacl.430/",
    doi = "10.18653/v1/2025.findings-naacl.430",
    pages = "7724--7743",
    ISBN = "979-8-89176-195-7"
}

@inproceedings{chaszczewicz-etal-2024-multi,
    title = "Multi-Level Feedback Generation with Large Language Models for Empowering Novice Peer Counselors",
    author = "Chaszczewicz, Alicja  and
      Shah, Raj  and
      Louie, Ryan  and
      Arnow, Bruce  and
      Kraut, Robert  and
      Yang, Diyi",
    editor = "Ku, Lun-Wei  and
      Martins, Andre  and
      Srikumar, Vivek",
    booktitle = "Proceedings of the 62nd Annual Meeting of the Association for Computational Linguistics (Volume 1: Long Papers)",
    month = aug,
    year = "2024",
    address = "Bangkok, Thailand",
    publisher = "Association for Computational Linguistics",
    url = "https://aclanthology.org/2024.acl-long.227/",
    doi = "10.18653/v1/2024.acl-long.227",
    pages = "4130--4161"
}

@inproceedings{tsakalidis-etal-2022-identifying,
    title = "Identifying Moments of Change from Longitudinal User Text",
    author = "Tsakalidis, Adam  and
      Nanni, Federico  and
      Hills, Anthony  and
      Chim, Jenny  and
      Song, Jiayu  and
      Liakata, Maria",
    editor = "Muresan, Smaranda  and
      Nakov, Preslav  and
      Villavicencio, Aline",
    booktitle = "Proceedings of the 60th Annual Meeting of the Association for Computational Linguistics (Volume 1: Long Papers)",
    month = may,
    year = "2022",
    address = "Dublin, Ireland",
    publisher = "Association for Computational Linguistics",
    url = "https://aclanthology.org/2022.acl-long.318/",
    doi = "10.18653/v1/2022.acl-long.318",
    pages = "4647--4660"
}

@inproceedings{juhng-etal-2023-discourse,
    title = "Discourse-Level Representations can Improve Prediction of Degree of Anxiety",
    author = "Juhng, Swanie  and
      Matero, Matthew  and
      Varadarajan, Vasudha  and
      Eichstaedt, Johannes  and
      V Ganesan, Adithya  and
      Schwartz, H. Andrew",
    editor = "Rogers, Anna  and
      Boyd-Graber, Jordan  and
      Okazaki, Naoaki",
    booktitle = "Proceedings of the 61st Annual Meeting of the Association for Computational Linguistics (Volume 2: Short Papers)",
    month = jul,
    year = "2023",
    address = "Toronto, Canada",
    publisher = "Association for Computational Linguistics",
    url = "https://aclanthology.org/2023.acl-short.128/",
    doi = "10.18653/v1/2023.acl-short.128",
    pages = "1500--1511"
}

@article{cruz2025artificial,
  title={Artificial intelligence in mental health care: a systematic review of diagnosis, monitoring, and intervention applications},
  author={Cruz-Gonzalez, Pablo and He, Aaron Wan-Jia and Lam, Elly PoPo and Ng, Ingrid Man Ching and Li, Mandy Wingman and Hou, Rangchun and Chan, Jackie Ngai-Man and Sahni, Yuvraj and Guasch, Nestor Vinas and Miller, Tiev and others},
  journal={Psychological medicine},
  volume={55},
  pages={e18},
  year={2025},
  publisher={Cambridge University Press},
url={https://www.cambridge.org/core/journals/psychological-medicine/article/artificial-intelligence-in-mental-health-care-a-systematic-review-of-diagnosis-monitoring-and-intervention-applications/04DBD2D05976C9B1873B475018695418}
}

@inproceedings{kumar-etal-2024-mental,
    title = "Mental Disorder Classification via Temporal Representation of Text",
    author = "Kumar, Raja  and
      Maharaj, Kishan  and
      Saxena, Ashita  and
      Bhattacharyya, Pushpak",
    editor = "Al-Onaizan, Yaser  and
      Bansal, Mohit  and
      Chen, Yun-Nung",
    booktitle = "Findings of the Association for Computational Linguistics: EMNLP 2024",
    month = nov,
    year = "2024",
    address = "Miami, Florida, USA",
    publisher = "Association for Computational Linguistics",
    url = "https://aclanthology.org/2024.findings-emnlp.639/",
    doi = "10.18653/v1/2024.findings-emnlp.639",
    pages = "10901--10916"
}

@inproceedings{liu2021towards,
  title={Towards Emotional Support Dialog Systems},
  author={Liu, Siyang and Zheng, Chujie and Demasi, Orianna and Sabour, Sahand and Li, Yu and Yu, Zhou and Jiang, Yong and Huang, Minlie},
  booktitle={Proceedings of the 59th Annual Meeting of the Association for Computational Linguistics and the 11th International Joint Conference on Natural Language Processing (Volume 1: Long Papers)},
  pages={3469--3483},
  year={2021},
url={https://aclanthology.org/2021.acl-long.269/}
}

@inproceedings{park2020suicidal,
    title = "Suicidal Risk Detection for Military Personnel",
    author = "Park, Sungjoon  and
      Park, Kiwoong  and
      Ahn, Jaimeen  and
      Oh, Alice",
    editor = "Webber, Bonnie  and
      Cohn, Trevor  and
      He, Yulan  and
      Liu, Yang",
    booktitle = "Proceedings of the 2020 Conference on Empirical Methods in Natural Language Processing (EMNLP)",
    month = nov,
    year = "2020",
    address = "Online",
    publisher = "Association for Computational Linguistics",
    url = "https://aclanthology.org/2020.emnlp-main.198/",
    doi = "10.18653/v1/2020.emnlp-main.198",
    pages = "2523--2531"
}

@inproceedings{zhang-etal-2025-speecht,
    title = "{S}peech{T}-{RAG}: Reliable Depression Detection in {LLM}s with Retrieval-Augmented Generation Using Speech Timing Information",
    author = "Zhang, Xiangyu  and
      Liu, Hexin  and
      Zhang, Qiquan  and
      Ahmed, Beena  and
      Epps, Julien",
    editor = "Che, Wanxiang  and
      Nabende, Joyce  and
      Shutova, Ekaterina  and
      Pilehvar, Mohammad Taher",
    booktitle = "Findings of the Association for Computational Linguistics: ACL 2025",
    month = jul,
    year = "2025",
    address = "Vienna, Austria",
    publisher = "Association for Computational Linguistics",
    url = "https://aclanthology.org/2025.findings-acl.521/",
    doi = "10.18653/v1/2025.findings-acl.521",
    pages = "10019--10030",
    ISBN = "979-8-89176-256-5"
}

@inproceedings{qiu-lan-2025-psydial,
    title = "{P}sy{D}ial: A Large-scale Long-term Conversational Dataset for Mental Health Support",
    author = "Qiu, Huachuan  and
      Lan, Zhenzhong",
    editor = "Che, Wanxiang  and
      Nabende, Joyce  and
      Shutova, Ekaterina  and
      Pilehvar, Mohammad Taher",
    booktitle = "Proceedings of the 63rd Annual Meeting of the Association for Computational Linguistics (Volume 1: Long Papers)",
    month = jul,
    year = "2025",
    address = "Vienna, Austria",
    publisher = "Association for Computational Linguistics",
    url = "https://aclanthology.org/2025.acl-long.1049/",
    doi = "10.18653/v1/2025.acl-long.1049",
    pages = "21624--21655",
    ISBN = "979-8-89176-251-0"
}

@article{cuijpers2019role,
  title={The role of common factors in psychotherapy outcomes},
  author={Cuijpers, Pim and Reijnders, Mirjam and Huibers, Marcus JH},
  journal={Annual review of clinical psychology},
  volume={15},
  number={1},
  pages={207--231},
  year={2019},
  publisher={Annual Reviews},
url={https://www.annualreviews.org/content/journals/10.1146/annurev-clinpsy-050718-095424}
}

@book{hofmann2013art,
  title={The art and science of psychotherapy},
  author={Hofmann, Stefan G and Weinberger, Joel},
  year={2013},
  publisher={Routledge},
url={https://api.taylorfrancis.com/content/books/mono/download?identifierName=doi&identifierValue=10.4324/9780203943427&type=googlepdf}
}

@article{torous2019creating,
  title={Creating a digital health smartphone app and digital phenotyping platform for mental health and diverse healthcare needs: an interdisciplinary and collaborative approach},
  author={Torous, John and Wisniewski, Hannah and Bird, Bruce and Carpenter, Elizabeth and David, Gary and Elejalde, Eduardo and Fulford, Dan and Guimond, Synthia and Hays, Ryan and Henson, Philip and others},
  journal={Journal of Technology in Behavioral Science},
  volume={4},
  number={2},
  pages={73--85},
  year={2019},
  publisher={Springer},
url={https://link.springer.com/article/10.1007/s41347-019-00095-w}
}

@article{kleinberg2016inherent,
  title={Inherent trade-offs in the fair determination of risk scores},
  author={Kleinberg, Jon and Mullainathan, Sendhil and Raghavan, Manish},
  journal={arXiv preprint arXiv:1609.05807},
  year={2016},
url={https://arxiv.org/abs/1609.05807}
}

@article{borsboom2017network,
  title={A network theory of mental disorders},
  author={Borsboom, Denny},
  journal={World psychiatry},
  volume={16},
  number={1},
  pages={5--13},
  year={2017},
  publisher={Wiley Online Library},
url={https://onlinelibrary.wiley.com/doi/full/10.1002/wps.20375}
}

@article{scheffer2024dynamical,
  title={A dynamical systems view of psychiatric disorders—theory: a review},
  author={Scheffer, Marten and Bockting, Claudi L and Borsboom, Denny and Cools, Roshan and Delecroix, Clara and Hartmann, Jessica A and Kendler, Kenneth S and van de Leemput, Ingrid and Van Der Maas, Han LJ and van Nes, Egbert and others},
  journal={JAMA psychiatry},
  volume={81},
  number={6},
  pages={618--623},
  year={2024},
  publisher={American Medical Association},
url={https://jamanetwork.com/journals/jamapsychiatry/article-abstract/2817087}
}

@misc{ong2025llm,
      title={Using Large Language Models to Create Personalized Networks From Therapy Sessions}, 
      author={Clarissa W. Ong and Hiba Arnaout and Kate Sheehan and Estella Fox and Eugen Owtscharow and Iryna Gurevych},
      year={2025},
      eprint={2512.05836},
      archivePrefix={arXiv},
      primaryClass={cs.AI},
      url={https://arxiv.org/abs/2512.05836}, 
}

@article{diagnostic2013statistical,
  title={Statistical manual of mental disorders: DSM-5 (ed.) Washington},
  author={Diagnostic, AP},
  journal={DC: American Psychiatric Association},
  year={2013}
}

@article{spitzer2006brief,
  title={A brief measure for assessing generalized anxiety disorder: the GAD-7},
  author={Spitzer, Robert L and Kroenke, Kurt and Williams, Janet BW and L{\"o}we, Bernd},
  journal={Archives of internal medicine},
  volume={166},
  number={10},
  pages={1092--1097},
  year={2006},
  publisher={American Medical Association},
url={https://jamanetwork.com/journals/jamainternalmedicine/fullarticle/410326}
}

@article{spitzer1999validation,
  title={Validation and utility of a self-report version of PRIME-MD: the PHQ primary care study},
  author={Spitzer, Robert L and Kroenke, Kurt and Williams, Janet BW and Patient Health Questionnaire Primary Care Study Group and others},
  journal={jama},
  volume={282},
  number={18},
  pages={1737--1744},
  year={1999},
  publisher={American Medical Association},
url={https://jamanetwork.com/journals/jama/fullarticle/192080}
}

@article{lavelle2025explainable,
  title={An explainable artificial intelligence handbook for psychologists: Methods, opportunities, and challenges.},
  author={Lavelle-Hill, Rosa and Smith, Gavin and Deininger, Hannah and Murayama, Kou},
  journal={Psychological Methods},
  year={2025},
  publisher={American Psychological Association},
url={https://psycnet.apa.org/fulltext/2026-46377-001.html}
}

@article{reddy2024generative,
  title={Generative AI in healthcare: an implementation science informed translational path on application, integration and governance},
  author={Reddy, Sandeep},
  journal={Implementation Science},
  volume={19},
  number={1},
  pages={27},
  year={2024},
  publisher={Springer},
url={https://link.springer.com/article/10.1186/s13012-024-01357-9}
}

@inproceedings{lyon2023bridging,
  title={Bridging HCI and implementation science for innovation adoption and public health impact},
  author={Lyon, Aaron and Munson, Sean A and Reddy, Madhu and Schueller, Stephen M and Agapie, Elena and Yarosh, Svetlana and Dopp, Alex and von Thiele Schwarz, Ulrica and Doherty, Gavin and Graham, Andrea K and others},
  booktitle={Extended Abstracts of the 2023 CHI Conference on Human Factors in Computing Systems},
  pages={1--7},
  year={2023},
url={https://dl.acm.org/doi/full/10.1145/3544549.3574132}
}

@book{meyer2010understanding,
  title={Understanding measurement: reliability},
  author={Meyer, Patrick},
  year={2010},
  publisher={Oxford University Press},
url={https://psycnet.apa.org/record/2010-09329-000}
}

@book{cohen1988psychological,
  title={Psychological testing: An introduction to tests \& measurement.},
  author={Cohen, Ronald Jay and Montague, Pamela and Nathanson, Linda Sue and Swerdlik, Mark E},
  year={1988},
  publisher={Mayfield Publishing Co},
url={https://psycnet.apa.org/record/1996-97180-000}
}

@book{reynolds2021mastering,
  title={Mastering modern psychological testing},
  author={Reynolds, Cecil R and Livingston, RA},
  year={2021},
  publisher={Springer},
url={https://link.springer.com/book/10.1007/978-3-030-59455-8}
}

@article{cook2006current,
  title={Current concepts in validity and reliability for psychometric instruments: theory and application},
  author={Cook, David A and Beckman, Thomas J},
  journal={The American journal of medicine},
  volume={119},
  number={2},
  pages={166--e7},
  year={2006},
  publisher={Elsevier},
url={https://www.sciencedirect.com/science/article/pii/S0002934305010375}
}

@inproceedings{atapattu-etal-2025-exploring,
    title = "Exploring the Role of Mental Health Conversational Agents in Training Medical Students and Professionals: A Systematic Literature Review",
    author = "Atapattu, Thushari  and
      Thilakaratne, Menasha  and
      Do, Duc Nhan  and
      Herath, Mahen  and
      Falkner, Katrina E.",
    editor = "Che, Wanxiang  and
      Nabende, Joyce  and
      Shutova, Ekaterina  and
      Pilehvar, Mohammad Taher",
    booktitle = "Findings of the Association for Computational Linguistics: ACL 2025",
    month = jul,
    year = "2025",
    address = "Vienna, Austria",
    publisher = "Association for Computational Linguistics",
    url = "https://aclanthology.org/2025.findings-acl.1069/",
    doi = "10.18653/v1/2025.findings-acl.1069",
    pages = "20785--20798",
    ISBN = "979-8-89176-256-5"
}

@inproceedings{sage-etal-2025-survey,
    title = "A Survey of Cognitive Distortion Detection and Classification in {NLP}",
    author = "Sage, Archie  and
      Keppens, Jeroen  and
      Yannakoudakis, Helen",
    editor = "Christodoulopoulos, Christos  and
      Chakraborty, Tanmoy  and
      Rose, Carolyn  and
      Peng, Violet",
    booktitle = "Findings of the Association for Computational Linguistics: EMNLP 2025",
    month = nov,
    year = "2025",
    address = "Suzhou, China",
    publisher = "Association for Computational Linguistics",
    url = "https://aclanthology.org/2025.findings-emnlp.804/",
    doi = "10.18653/v1/2025.findings-emnlp.804",
    pages = "14884--14899",
    ISBN = "979-8-89176-335-7"
}

@inproceedings{na-etal-2025-survey,
    title = "A Survey of Large Language Models in Psychotherapy: Current Landscape and Future Directions",
    author = "Na, Hongbin  and
      Hua, Yining  and
      Wang, Zimu  and
      Shen, Tao  and
      Yu, Beibei  and
      Wang, Lilin  and
      Wang, Wei  and
      Torous, John  and
      Chen, Ling",
    editor = "Che, Wanxiang  and
      Nabende, Joyce  and
      Shutova, Ekaterina  and
      Pilehvar, Mohammad Taher",
    booktitle = "Findings of the Association for Computational Linguistics: ACL 2025",
    month = jul,
    year = "2025",
    address = "Vienna, Austria",
    publisher = "Association for Computational Linguistics",
    url = "https://aclanthology.org/2025.findings-acl.385/",
    doi = "10.18653/v1/2025.findings-acl.385",
    pages = "7362--7376",
    ISBN = "979-8-89176-256-5"
}

@inproceedings{sathvik-etal-2025-help,
    title = "{M}-Help: Using Social Media Data to Detect Mental Health Help-Seeking Signals",
    author = "Sathvik, Msvpj  and
      Shaik, Zuhair Hasan  and
      Gupta, Vivek",
    editor = "Christodoulopoulos, Christos  and
      Chakraborty, Tanmoy  and
      Rose, Carolyn  and
      Peng, Violet",
    booktitle = "Findings of the Association for Computational Linguistics: EMNLP 2025",
    month = nov,
    year = "2025",
    address = "Suzhou, China",
    publisher = "Association for Computational Linguistics",
    url = "https://aclanthology.org/2025.findings-emnlp.1225/",
    doi = "10.18653/v1/2025.findings-emnlp.1225",
    pages = "22510--22520",
    ISBN = "979-8-89176-335-7"
}

@inproceedings{kuzmin-etal-2025-exploring,
    title = "Exploring Large Language Models for Detecting Mental Disorders",
    author = "Kuzmin, Gleb  and
      Strepetov, Petr  and
      Stankevich, Maksim  and
      Chudova, Natalia  and
      Shelmanov, Artem  and
      Smirnov, Ivan",
    editor = "Christodoulopoulos, Christos  and
      Chakraborty, Tanmoy  and
      Rose, Carolyn  and
      Peng, Violet",
    booktitle = "Proceedings of the 2025 Conference on Empirical Methods in Natural Language Processing",
    month = nov,
    year = "2025",
    address = "Suzhou, China",
    publisher = "Association for Computational Linguistics",
    url = "https://aclanthology.org/2025.emnlp-main.1752/",
    doi = "10.18653/v1/2025.emnlp-main.1752",
    pages = "34523--34547",
    ISBN = "979-8-89176-332-6"
}

@inproceedings{kim-kim-2025-koacd,
    title = "{K}o{ACD}: The First {K}orean Adolescent Dataset for Cognitive Distortion Analysis via Role-Switching Multi-{LLM} Negotiation",
    author = "Kim, Jun Seo  and
      Kim, Hye Hyeon",
    editor = "Christodoulopoulos, Christos  and
      Chakraborty, Tanmoy  and
      Rose, Carolyn  and
      Peng, Violet",
    booktitle = "Findings of the Association for Computational Linguistics: EMNLP 2025",
    month = nov,
    year = "2025",
    address = "Suzhou, China",
    publisher = "Association for Computational Linguistics",
    url = "https://aclanthology.org/2025.findings-emnlp.1202/",
    doi = "10.18653/v1/2025.findings-emnlp.1202",
    pages = "22050--22078",
    ISBN = "979-8-89176-335-7"
}

@inproceedings{bn-etal-2025-real,
    title = "How Real Are Synthetic Therapy Conversations? Evaluating Fidelity in Prolonged Exposure Dialogues",
    author = "Bn, Suhas  and
      Mattioli, Dominik O.  and
      Sherrill, Andrew M.  and
      Arriaga, Rosa I.  and
      Wiese, Christopher  and
      Abdullah, Saeed",
    editor = "Christodoulopoulos, Christos  and
      Chakraborty, Tanmoy  and
      Rose, Carolyn  and
      Peng, Violet",
    booktitle = "Findings of the Association for Computational Linguistics: EMNLP 2025",
    month = nov,
    year = "2025",
    address = "Suzhou, China",
    publisher = "Association for Computational Linguistics",
    url = "https://aclanthology.org/2025.findings-emnlp.1144/",
    doi = "10.18653/v1/2025.findings-emnlp.1144",
    pages = "20986--20995",
    ISBN = "979-8-89176-335-7"
}

@inproceedings{reuben-etal-2025-assessment,
    title = "Assessment and manipulation of latent constructs in pre-trained language models using psychometric scales",
    author = "Reuben, Maor  and
      Slobodin, Ortal  and
      Cohen, Idan-Chaim  and
      Elyashar, Aviad  and
      Braun-Lewensohn, Orna  and
      Cohen, Odeya  and
      Puzis, Rami",
    editor = "Che, Wanxiang  and
      Nabende, Joyce  and
      Shutova, Ekaterina  and
      Pilehvar, Mohammad Taher",
    booktitle = "Proceedings of the 63rd Annual Meeting of the Association for Computational Linguistics (Volume 1: Long Papers)",
    month = jul,
    year = "2025",
    address = "Vienna, Austria",
    publisher = "Association for Computational Linguistics",
    url = "https://aclanthology.org/2025.acl-long.121/",
    doi = "10.18653/v1/2025.acl-long.121",
    pages = "2433--2444",
    ISBN = "979-8-89176-251-0"
}

@inproceedings{chen-etal-2025-catch,
    title = "{CATCH}: A Novel Data Synthesis Framework for High Therapy Fidelity and Memory-Driven Planning Chain of Thought in {AI} Counseling",
    author = "Chen, Mingyu  and
      Lin, Jingkai  and
      Chu, Zhaojie  and
      Xing, Xiaofen  and
      Chen, Yirong  and
      Xu, Xiangmin",
    editor = "Christodoulopoulos, Christos  and
      Chakraborty, Tanmoy  and
      Rose, Carolyn  and
      Peng, Violet",
    booktitle = "Findings of the Association for Computational Linguistics: EMNLP 2025",
    month = nov,
    year = "2025",
    address = "Suzhou, China",
    publisher = "Association for Computational Linguistics",
    url = "https://aclanthology.org/2025.findings-emnlp.543/",
    doi = "10.18653/v1/2025.findings-emnlp.543",
    pages = "10254--10286",
    ISBN = "979-8-89176-335-7"
}

@inproceedings{hong-etal-2025-third,
    title = "Third-Person Appraisal Agent: Simulating Human Emotional Reasoning in Text with Large Language Models",
    author = "Hong, Simin  and
      Sun, Jun  and
      Chen, Hongyang",
    editor = "Christodoulopoulos, Christos  and
      Chakraborty, Tanmoy  and
      Rose, Carolyn  and
      Peng, Violet",
    booktitle = "Findings of the Association for Computational Linguistics: EMNLP 2025",
    month = nov,
    year = "2025",
    address = "Suzhou, China",
    publisher = "Association for Computational Linguistics",
    url = "https://aclanthology.org/2025.findings-emnlp.1288/",
    doi = "10.18653/v1/2025.findings-emnlp.1288",
    pages = "23684--23701",
    ISBN = "979-8-89176-335-7"
}

@inproceedings{xie-etal-2025-psydt,
    title = "{P}sy{DT}: Using {LLM}s to Construct the Digital Twin of Psychological Counselor with Personalized Counseling Style for Psychological Counseling",
    author = "Xie, Haojie  and
      Chen, Yirong  and
      Xing, Xiaofen  and
      Lin, Jingkai  and
      Xu, Xiangmin",
    editor = "Che, Wanxiang  and
      Nabende, Joyce  and
      Shutova, Ekaterina  and
      Pilehvar, Mohammad Taher",
    booktitle = "Proceedings of the 63rd Annual Meeting of the Association for Computational Linguistics (Volume 1: Long Papers)",
    month = jul,
    year = "2025",
    address = "Vienna, Austria",
    publisher = "Association for Computational Linguistics",
    url = "https://aclanthology.org/2025.acl-long.55/",
    doi = "10.18653/v1/2025.acl-long.55",
    pages = "1081--1115",
    ISBN = "979-8-89176-251-0"
}

@inproceedings{chen-etal-2025-mind,
    title = "{MIND}: Towards Immersive Psychological Healing with Multi-Agent Inner Dialogue",
    author = "Chen, Yujia  and
      Li, Changsong  and
      Wang, Yiming  and
      Ju, Tianjie  and
      Xiao, Qingqing  and
      Zhang, Nan  and
      Kong, Zifan  and
      Wang, Peng  and
      Yan, Binyu",
    editor = "Christodoulopoulos, Christos  and
      Chakraborty, Tanmoy  and
      Rose, Carolyn  and
      Peng, Violet",
    booktitle = "Findings of the Association for Computational Linguistics: EMNLP 2025",
    month = nov,
    year = "2025",
    address = "Suzhou, China",
    publisher = "Association for Computational Linguistics",
    url = "https://aclanthology.org/2025.findings-emnlp.499/",
    doi = "10.18653/v1/2025.findings-emnlp.499",
    pages = "9380--9413",
    ISBN = "979-8-89176-335-7"
}

@inproceedings{qiu-etal-2025-emoagent,
    title = "{E}mo{A}gent: Assessing and Safeguarding Human-{AI} Interaction for Mental Health Safety",
    author = "Qiu, Jiahao  and
      He, Yinghui  and
      Juan, Xinzhe  and
      Wang, Yimin  and
      Liu, Yuhan  and
      Yao, Zixin  and
      Wu, Yue  and
      Jiang, Xun  and
      Yang, Ling  and
      Wang, Mengdi",
    editor = "Christodoulopoulos, Christos  and
      Chakraborty, Tanmoy  and
      Rose, Carolyn  and
      Peng, Violet",
    booktitle = "Proceedings of the 2025 Conference on Empirical Methods in Natural Language Processing",
    month = nov,
    year = "2025",
    address = "Suzhou, China",
    publisher = "Association for Computational Linguistics",
    url = "https://aclanthology.org/2025.emnlp-main.594/",
    doi = "10.18653/v1/2025.emnlp-main.594",
    pages = "11752--11767",
    ISBN = "979-8-89176-332-6"
}

@inproceedings{zheng-etal-2025-promind,
    title = "{P}ro{M}ind-{LLM}: Proactive Mental Health Care via Causal Reasoning with Sensor Data",
    author = "Zheng, Xinzhe  and
      Ji, Sijie  and
      Sun, Jiawei  and
      Chen, Renqi  and
      Gao, Wei  and
      Srivastava, Mani",
    editor = "Che, Wanxiang  and
      Nabende, Joyce  and
      Shutova, Ekaterina  and
      Pilehvar, Mohammad Taher",
    booktitle = "Findings of the Association for Computational Linguistics: ACL 2025",
    month = jul,
    year = "2025",
    address = "Vienna, Austria",
    publisher = "Association for Computational Linguistics",
    url = "https://aclanthology.org/2025.findings-acl.1033/",
    doi = "10.18653/v1/2025.findings-acl.1033",
    pages = "20150--20171",
    ISBN = "979-8-89176-256-5"
}

@inproceedings{zhou-etal-2025-crisp,
    title = "Crisp: Cognitive Restructuring of Negative Thoughts through Multi-turn Supportive Dialogues",
    author = "Zhou, Jinfeng  and
      Chen, Yuxuan  and
      Yin, Jianing  and
      Huang, Yongkang  and
      Shi, Yihan  and
      Zhang, Xikun  and
      Peng, Libiao  and
      Zhang, Rongsheng  and
      Lv, Tangjie  and
      Hu, Zhipeng  and
      Wang, Hongning  and
      Huang, Minlie",
    editor = "Christodoulopoulos, Christos  and
      Chakraborty, Tanmoy  and
      Rose, Carolyn  and
      Peng, Violet",
    booktitle = "Proceedings of the 2025 Conference on Empirical Methods in Natural Language Processing",
    month = nov,
    year = "2025",
    address = "Suzhou, China",
    publisher = "Association for Computational Linguistics",
    url = "https://aclanthology.org/2025.emnlp-main.1652/",
    doi = "10.18653/v1/2025.emnlp-main.1652",
    pages = "32462--32491",
    ISBN = "979-8-89176-332-6"
}

@inproceedings{song-etal-2025-rationale,
    title = "Does Rationale Quality Matter? Enhancing Mental Disorder Detection via Selective Reasoning Distillation",
    author = "Song, Hoyun  and
      Lee, Huije  and
      Shin, Jisu  and
      Cho, Sukmin  and
      Ko, Changgeon  and
      Park, Jong C.",
    editor = "Che, Wanxiang  and
      Nabende, Joyce  and
      Shutova, Ekaterina  and
      Pilehvar, Mohammad Taher",
    booktitle = "Findings of the Association for Computational Linguistics: ACL 2025",
    month = jul,
    year = "2025",
    address = "Vienna, Austria",
    publisher = "Association for Computational Linguistics",
    url = "https://aclanthology.org/2025.findings-acl.1119/",
    doi = "10.18653/v1/2025.findings-acl.1119",
    pages = "21738--21756",
    ISBN = "979-8-89176-256-5"
}

@inproceedings{yang-etal-2025-consistent,
    title = "Consistent Client Simulation for Motivational Interviewing-based Counseling",
    author = "Yang, Yizhe  and
      Achananuparp, Palakorn  and
      Huang, Heyan  and
      Jiang, Jing  and
      Lim, Nicholas Gabriel  and
      Ern, Cameron Tan Shi  and
      Kit, Phey Ling  and
      Xiuhui, Jenny Giam  and
      Pinto, John  and
      Lim, Ee-Peng",
    editor = "Che, Wanxiang  and
      Nabende, Joyce  and
      Shutova, Ekaterina  and
      Pilehvar, Mohammad Taher",
    booktitle = "Proceedings of the 63rd Annual Meeting of the Association for Computational Linguistics (Volume 1: Long Papers)",
    month = jul,
    year = "2025",
    address = "Vienna, Austria",
    publisher = "Association for Computational Linguistics",
    url = "https://aclanthology.org/2025.acl-long.1021/",
    doi = "10.18653/v1/2025.acl-long.1021",
    pages = "20959--20998",
    ISBN = "979-8-89176-251-0"
}

@inproceedings{yang-etal-2025-cami,
    title = "{CAMI}: A Counselor Agent Supporting Motivational Interviewing through State Inference and Topic Exploration",
    author = "Yang, Yizhe  and
      Achananuparp, Palakorn  and
      Huang, Heyan  and
      Jiang, Jing  and
      Kit, Phey Ling  and
      Lim, Nicholas Gabriel  and
      Ern, Cameron Tan Shi  and
      Lim, Ee-Peng",
    editor = "Che, Wanxiang  and
      Nabende, Joyce  and
      Shutova, Ekaterina  and
      Pilehvar, Mohammad Taher",
    booktitle = "Proceedings of the 63rd Annual Meeting of the Association for Computational Linguistics (Volume 1: Long Papers)",
    month = jul,
    year = "2025",
    address = "Vienna, Austria",
    publisher = "Association for Computational Linguistics",
    url = "https://aclanthology.org/2025.acl-long.1024/",
    doi = "10.18653/v1/2025.acl-long.1024",
    pages = "21037--21081",
    ISBN = "979-8-89176-251-0"
}

@inproceedings{wang-etal-2025-feel,
    title = "Feel the Difference? A Comparative Analysis of Emotional Arcs in Real and {LLM}-Generated {CBT} Sessions",
    author = "Wang, Xiaoyi  and
      Zhang, Jiwei  and
      Zhang, Guangtao  and
      Guo, Honglei",
    editor = "Christodoulopoulos, Christos  and
      Chakraborty, Tanmoy  and
      Rose, Carolyn  and
      Peng, Violet",
    booktitle = "Findings of the Association for Computational Linguistics: EMNLP 2025",
    month = nov,
    year = "2025",
    address = "Suzhou, China",
    publisher = "Association for Computational Linguistics",
    url = "https://aclanthology.org/2025.findings-emnlp.1089/",
    doi = "10.18653/v1/2025.findings-emnlp.1089",
    pages = "19999--20017",
    ISBN = "979-8-89176-335-7"
}

@inproceedings{aghakhani-etal-2025-conversation,
    title = "From Conversation to Automation: Leveraging {LLM}s for Problem-Solving Therapy Analysis",
    author = "Aghakhani, Elham  and
      Wang, Lu  and
      Washington, Karla T.  and
      Demiris, George  and
      Huh-Yoo, Jina  and
      Rezapour, Rezvaneh",
    editor = "Che, Wanxiang  and
      Nabende, Joyce  and
      Shutova, Ekaterina  and
      Pilehvar, Mohammad Taher",
    booktitle = "Findings of the Association for Computational Linguistics: ACL 2025",
    month = jul,
    year = "2025",
    address = "Vienna, Austria",
    publisher = "Association for Computational Linguistics",
    url = "https://aclanthology.org/2025.findings-acl.1292/",
    doi = "10.18653/v1/2025.findings-acl.1292",
    pages = "25189--25207",
    ISBN = "979-8-89176-256-5"
}

@inproceedings{feng-etal-2025-reframe,
    title = "Reframe Your Life Story: Interactive Narrative Therapist and Innovative Moment Assessment with Large Language Models",
    author = "Feng, Yi  and
      Wang, Jiaqi  and
      Zhang, Wenxuan  and
      Chen, Zhuang  and
      Yutong, Shen  and
      Xiao, Xiyao  and
      Huang, Minlie  and
      Jing, Liping  and
      Yu, Jian",
    editor = "Christodoulopoulos, Christos  and
      Chakraborty, Tanmoy  and
      Rose, Carolyn  and
      Peng, Violet",
    booktitle = "Proceedings of the 2025 Conference on Empirical Methods in Natural Language Processing",
    month = nov,
    year = "2025",
    address = "Suzhou, China",
    publisher = "Association for Computational Linguistics",
    url = "https://aclanthology.org/2025.emnlp-main.1245/",
    doi = "10.18653/v1/2025.emnlp-main.1245",
    pages = "24495--24520",
    ISBN = "979-8-89176-332-6"
}

@inproceedings{lee-etal-2025-heart,
    title = "From Heart to Words: Generating Empathetic Responses via Integrated Figurative Language and Semantic Context Signals",
    author = "Lee, Gyeongeun  and
      Wang, Zhu  and
      Ravi, Sathya N.  and
      Parde, Natalie",
    editor = "Che, Wanxiang  and
      Nabende, Joyce  and
      Shutova, Ekaterina  and
      Pilehvar, Mohammad Taher",
    booktitle = "Findings of the Association for Computational Linguistics: ACL 2025",
    month = jul,
    year = "2025",
    address = "Vienna, Austria",
    publisher = "Association for Computational Linguistics",
    url = "https://aclanthology.org/2025.findings-acl.231/",
    doi = "10.18653/v1/2025.findings-acl.231",
    pages = "4490--4502",
    ISBN = "979-8-89176-256-5"
}

@inproceedings{zhang-etal-2025-explainable,
    title = "Explainable Depression Detection in Clinical Interviews with Personalized Retrieval-Augmented Generation",
    author = "Zhang, Linhai  and
      Gao, Ziyang  and
      Zhou, Deyu  and
      He, Yulan",
    editor = "Che, Wanxiang  and
      Nabende, Joyce  and
      Shutova, Ekaterina  and
      Pilehvar, Mohammad Taher",
    booktitle = "Findings of the Association for Computational Linguistics: ACL 2025",
    month = jul,
    year = "2025",
    address = "Vienna, Austria",
    publisher = "Association for Computational Linguistics",
    url = "https://aclanthology.org/2025.findings-acl.517/",
    doi = "10.18653/v1/2025.findings-acl.517",
    pages = "9927--9944",
    ISBN = "979-8-89176-256-5"
}

@inproceedings{zhang-poellabauer-2025-mitigating,
    title = "Mitigating Interviewer Bias in Multimodal Depression Detection: An Approach with Adversarial Learning and Contextual Positional Encoding",
    author = "Zhang, Enshi  and
      Poellabauer, Christian",
    editor = "Christodoulopoulos, Christos  and
      Chakraborty, Tanmoy  and
      Rose, Carolyn  and
      Peng, Violet",
    booktitle = "Findings of the Association for Computational Linguistics: EMNLP 2025",
    month = nov,
    year = "2025",
    address = "Suzhou, China",
    publisher = "Association for Computational Linguistics",
    url = "https://aclanthology.org/2025.findings-emnlp.650/",
    doi = "10.18653/v1/2025.findings-emnlp.650",
    pages = "12169--12188",
    ISBN = "979-8-89176-335-7"
}

@inproceedings{haydarov-etal-2025-towards,
    title = "Towards {AI}-Assisted Psychotherapy: Emotion-Guided Generative Interventions",
    author = "Haydarov, Kilichbek  and
      Mohamed, Youssef  and
      Goldenhersch, Emilio  and
      OCallaghan, Paul  and
      Li, Li-jia  and
      Elhoseiny, Mohamed",
    editor = "Christodoulopoulos, Christos  and
      Chakraborty, Tanmoy  and
      Rose, Carolyn  and
      Peng, Violet",
    booktitle = "Proceedings of the 2025 Conference on Empirical Methods in Natural Language Processing",
    month = nov,
    year = "2025",
    address = "Suzhou, China",
    publisher = "Association for Computational Linguistics",
    url = "https://aclanthology.org/2025.emnlp-main.1664/",
    doi = "10.18653/v1/2025.emnlp-main.1664",
    pages = "32724--32743",
    ISBN = "979-8-89176-332-6"
}

@inproceedings{zhai-etal-2025-mentalglm,
    title = "{M}ental{GLM} Series: Explainable Large Language Models for Mental Health Analysis on {C}hinese Social Media",
    author = "Zhai, Wei  and
      Bai, Nan  and
      Zhao, Qing  and
      Li, Jianqiang  and
      Wang, Fan  and
      Qi, Hongzhi  and
      Jiang, Meng  and
      Wang, Xiaoqin  and
      Yang, Bing Xiang  and
      Fu, Guanghui",
    editor = "Christodoulopoulos, Christos  and
      Chakraborty, Tanmoy  and
      Rose, Carolyn  and
      Peng, Violet",
    booktitle = "Proceedings of the 2025 Conference on Empirical Methods in Natural Language Processing",
    month = nov,
    year = "2025",
    address = "Suzhou, China",
    publisher = "Association for Computational Linguistics",
    url = "https://aclanthology.org/2025.emnlp-main.686/",
    doi = "10.18653/v1/2025.emnlp-main.686",
    pages = "13599--13614",
    ISBN = "979-8-89176-332-6"
}

@inproceedings{agarwal-etal-2025-redepress,
    title = "{R}e{D}epress: A Cognitive Framework for Detecting Depression Relapse from Social Media",
    author = "Agarwal, Aakash Kumar  and
      Bhattacharjee, Saprativa  and
      Rastogi, Mauli  and
      Jacob, Jemima S.  and
      Banerjee, Biplab  and
      Gupta, Rashmi  and
      Bhattacharyya, Pushpak",
    editor = "Christodoulopoulos, Christos  and
      Chakraborty, Tanmoy  and
      Rose, Carolyn  and
      Peng, Violet",
    booktitle = "Proceedings of the 2025 Conference on Empirical Methods in Natural Language Processing",
    month = nov,
    year = "2025",
    address = "Suzhou, China",
    publisher = "Association for Computational Linguistics",
    url = "https://aclanthology.org/2025.emnlp-main.1758/",
    doi = "10.18653/v1/2025.emnlp-main.1758",
    pages = "34652--34670",
    ISBN = "979-8-89176-332-6"
}

@inproceedings{ravenda-etal-2025-llms,
    title = "Are {LLM}s effective psychological assessors? Leveraging adaptive {RAG} for interpretable mental health screening through psychometric practice",
    author = "Ravenda, Federico  and
      Bahrainian, Seyed Ali  and
      Raballo, Andrea  and
      Mira, Antonietta  and
      Kando, Noriko",
    editor = "Che, Wanxiang  and
      Nabende, Joyce  and
      Shutova, Ekaterina  and
      Pilehvar, Mohammad Taher",
    booktitle = "Proceedings of the 63rd Annual Meeting of the Association for Computational Linguistics (Volume 1: Long Papers)",
    month = jul,
    year = "2025",
    address = "Vienna, Austria",
    publisher = "Association for Computational Linguistics",
    url = "https://aclanthology.org/2025.acl-long.440/",
    doi = "10.18653/v1/2025.acl-long.440",
    pages = "8975--8991",
    ISBN = "979-8-89176-251-0"
}

@inproceedings{kim-etal-2025-mirror,
    title = "{MIRROR}: Multimodal Cognitive Reframing Therapy for Rolling with Resistance",
    author = "Kim, Subin  and
      Kim, Hoonrae  and
      Lee, Jihyun  and
      Jeon, Yejin  and
      Lee, Gary",
    editor = "Christodoulopoulos, Christos  and
      Chakraborty, Tanmoy  and
      Rose, Carolyn  and
      Peng, Violet",
    booktitle = "Proceedings of the 2025 Conference on Empirical Methods in Natural Language Processing",
    month = nov,
    year = "2025",
    address = "Suzhou, China",
    publisher = "Association for Computational Linguistics",
    url = "https://aclanthology.org/2025.emnlp-main.751/",
    doi = "10.18653/v1/2025.emnlp-main.751",
    pages = "14851--14880",
    ISBN = "979-8-89176-332-6"
}

@inproceedings{song-etal-2025-temporal,
    title = "Temporal reasoning for timeline summarisation in social media",
    author = "Song, Jiayu  and
      Akhter, Mahmud Elahi  and
      Atzil-Slonim, Dana  and
      Liakata, Maria",
    editor = "Che, Wanxiang  and
      Nabende, Joyce  and
      Shutova, Ekaterina  and
      Pilehvar, Mohammad Taher",
    booktitle = "Proceedings of the 63rd Annual Meeting of the Association for Computational Linguistics (Volume 1: Long Papers)",
    month = jul,
    year = "2025",
    address = "Vienna, Austria",
    publisher = "Association for Computational Linguistics",
    url = "https://aclanthology.org/2025.acl-long.1362/",
    doi = "10.18653/v1/2025.acl-long.1362",
    pages = "28085--28101",
    ISBN = "979-8-89176-251-0"
}

@inproceedings{ghosh-etal-2025-just,
    title = "Just a Scratch: Enhancing {LLM} Capabilities for Self-harm Detection through Intent Differentiation and Emoji Interpretation",
    author = "Ghosh, Soumitra  and
      Singh, Gopendra Vikram  and
      Shambhavi, Shambhavi  and
      Choudhury, Sabarna  and
      Ekbal, Asif",
    editor = "Che, Wanxiang  and
      Nabende, Joyce  and
      Shutova, Ekaterina  and
      Pilehvar, Mohammad Taher",
    booktitle = "Proceedings of the 63rd Annual Meeting of the Association for Computational Linguistics (Volume 1: Long Papers)",
    month = jul,
    year = "2025",
    address = "Vienna, Austria",
    publisher = "Association for Computational Linguistics",
    url = "https://aclanthology.org/2025.acl-long.1330/",
    doi = "10.18653/v1/2025.acl-long.1330",
    pages = "27428--27445",
    ISBN = "979-8-89176-251-0"
}

@inproceedings{bn-etal-2025-pursuit,
    title = "The Pursuit of Empathy: Evaluating Small Language Models for {PTSD} Dialogue Support",
    author = "Bn, Suhas  and
      Mahajan, Yash  and
      Mattioli, Dominik O.  and
      Sherrill, Andrew M.  and
      Arriaga, Rosa I.  and
      Wiese, Christopher  and
      Abdullah, Saeed",
    editor = "Christodoulopoulos, Christos  and
      Chakraborty, Tanmoy  and
      Rose, Carolyn  and
      Peng, Violet",
    booktitle = "Proceedings of the 2025 Conference on Empirical Methods in Natural Language Processing",
    month = nov,
    year = "2025",
    address = "Suzhou, China",
    publisher = "Association for Computational Linguistics",
    url = "https://aclanthology.org/2025.emnlp-main.1573/",
    doi = "10.18653/v1/2025.emnlp-main.1573",
    pages = "30888--30910",
    ISBN = "979-8-89176-332-6"
}

@inproceedings{singh-etal-2025-systematic,
    title = "Systematic Evaluation of Auto-Encoding and Large Language Model Representations for Capturing Author States and Traits",
    author = "Singh, Khushboo  and
      Varadarajan, Vasudha  and
      V. Ganesan, Adithya  and
      Nilsson, August H{\r{a}}kan  and
      Soni, Nikita  and
      Mahwish, Syeda  and
      Chitale, Pranav  and
      Boyd, Ryan L.  and
      Ungar, Lyle  and
      Rosenthal, Richard N.  and
      Schwartz, H. Andrew",
    editor = "Che, Wanxiang  and
      Nabende, Joyce  and
      Shutova, Ekaterina  and
      Pilehvar, Mohammad Taher",
    booktitle = "Findings of the Association for Computational Linguistics: ACL 2025",
    month = jul,
    year = "2025",
    address = "Vienna, Austria",
    publisher = "Association for Computational Linguistics",
    url = "https://aclanthology.org/2025.findings-acl.971/",
    doi = "10.18653/v1/2025.findings-acl.971",
    pages = "18955--18973",
    ISBN = "979-8-89176-256-5"
}

@inproceedings{bi-etal-2025-magi,
    title = "{MAGI}: Multi-Agent Guided Interview for Psychiatric Assessment",
    author = "Bi, Guanqun  and
      Chen, Zhuang  and
      Liu, Zhoufu  and
      Wang, Hongkai  and
      Xiao, Xiyao  and
      Xie, Yuqiang  and
      Zhang, Wen  and
      Huang, Yongkang  and
      Chen, Yuxuan  and
      Peng, Libiao  and
      Huang, Minlie",
    editor = "Che, Wanxiang  and
      Nabende, Joyce  and
      Shutova, Ekaterina  and
      Pilehvar, Mohammad Taher",
    booktitle = "Findings of the Association for Computational Linguistics: ACL 2025",
    month = jul,
    year = "2025",
    address = "Vienna, Austria",
    publisher = "Association for Computational Linguistics",
    url = "https://aclanthology.org/2025.findings-acl.1278/",
    doi = "10.18653/v1/2025.findings-acl.1278",
    pages = "24898--24921",
    ISBN = "979-8-89176-256-5"
}

@inproceedings{xu-etal-2025-multiagentesc,
    title = "{M}ulti{A}gent{ESC}: A {LLM}-based Multi-Agent Collaboration Framework for Emotional Support Conversation",
    author = "Xu, Yangyang  and
      Hu, Jinpeng  and
      Zhao, Zhuoer  and
      Duan, Zhangling  and
      Sun, Xiao  and
      Yang, Xun",
    editor = "Christodoulopoulos, Christos  and
      Chakraborty, Tanmoy  and
      Rose, Carolyn  and
      Peng, Violet",
    booktitle = "Proceedings of the 2025 Conference on Empirical Methods in Natural Language Processing",
    month = nov,
    year = "2025",
    address = "Suzhou, China",
    publisher = "Association for Computational Linguistics",
    url = "https://aclanthology.org/2025.emnlp-main.232/",
    doi = "10.18653/v1/2025.emnlp-main.232",
    pages = "4665--4681",
    ISBN = "979-8-89176-332-6"
}

@inproceedings{kim-etal-2025-dialogue,
    title = "Dialogue Systems for Emotional Support via Value Reinforcement",
    author = "Kim, Juhee  and
      Mok, Chunghu  and
      Lee, Jisun  and
      Kim, Hyang Sook  and
      Jo, Yohan",
    editor = "Che, Wanxiang  and
      Nabende, Joyce  and
      Shutova, Ekaterina  and
      Pilehvar, Mohammad Taher",
    booktitle = "Proceedings of the 63rd Annual Meeting of the Association for Computational Linguistics (Volume 1: Long Papers)",
    month = jul,
    year = "2025",
    address = "Vienna, Austria",
    publisher = "Association for Computational Linguistics",
    url = "https://aclanthology.org/2025.acl-long.1395/",
    doi = "10.18653/v1/2025.acl-long.1395",
    pages = "28733--28766",
    ISBN = "979-8-89176-251-0"
}

@inproceedings{liu-etal-2025-eeyore,
    title = "Eeyore: Realistic Depression Simulation via Expert-in-the-Loop Supervised and Preference Optimization",
    author = "Liu, Siyang  and
      Brie, Bianca  and
      Li, Wenda  and
      Biester, Laura  and
      Lee, Andrew  and
      Pennebaker, James  and
      Mihalcea, Rada",
    editor = "Che, Wanxiang  and
      Nabende, Joyce  and
      Shutova, Ekaterina  and
      Pilehvar, Mohammad Taher",
    booktitle = "Findings of the Association for Computational Linguistics: ACL 2025",
    month = jul,
    year = "2025",
    address = "Vienna, Austria",
    publisher = "Association for Computational Linguistics",
    url = "https://aclanthology.org/2025.findings-acl.707/",
    doi = "10.18653/v1/2025.findings-acl.707",
    pages = "13750--13770",
    ISBN = "979-8-89176-256-5"
}

@inproceedings{li-etal-2025-large-language-models-identify,
    title = "Can Large Language Models Identify Implicit Suicidal Ideation? An Empirical Evaluation",
    author = "Li, Tong  and
      Yang, Shu  and
      Wu, Junchao  and
      Wei, Jiyao  and
      Hu, Lijie  and
      Li, Mengdi  and
      Wong, Derek F.  and
      Oltmanns, Joshua R.  and
      Wang, Di",
    editor = "Christodoulopoulos, Christos  and
      Chakraborty, Tanmoy  and
      Rose, Carolyn  and
      Peng, Violet",
    booktitle = "Findings of the Association for Computational Linguistics: EMNLP 2025",
    month = nov,
    year = "2025",
    address = "Suzhou, China",
    publisher = "Association for Computational Linguistics",
    url = "https://aclanthology.org/2025.findings-emnlp.998/",
    doi = "10.18653/v1/2025.findings-emnlp.998",
    pages = "18392--18413",
    ISBN = "979-8-89176-335-7"
}

@inproceedings{lv-etal-2025-tracking,
    title = "Tracking Life{'}s Ups and Downs: Mining Life Events from Social Media Posts for Mental Health Analysis",
    author = "Lv, Minghao  and
      Chen, Siyuan  and
      Jin, Haoan  and
      Yuan, Minghao  and
      Ju, Qianqian  and
      Peng, Yujia  and
      Zhu, Kenny Q.  and
      Wu, Mengyue",
    editor = "Che, Wanxiang  and
      Nabende, Joyce  and
      Shutova, Ekaterina  and
      Pilehvar, Mohammad Taher",
    booktitle = "Proceedings of the 63rd Annual Meeting of the Association for Computational Linguistics (Volume 1: Long Papers)",
    month = jul,
    year = "2025",
    address = "Vienna, Austria",
    publisher = "Association for Computational Linguistics",
    url = "https://aclanthology.org/2025.acl-long.345/",
    doi = "10.18653/v1/2025.acl-long.345",
    pages = "6950--6965",
    ISBN = "979-8-89176-251-0"
}

@inproceedings{wang-etal-2025-annaagent,
    title = "{A}nna{A}gent: Dynamic Evolution Agent System with Multi-Session Memory for Realistic Seeker Simulation",
    author = "Wang, Ming  and
      Wang, Peidong  and
      Wu, Lin  and
      Yang, Xiaocui  and
      Wang, Daling  and
      Feng, Shi  and
      Chen, Yuxin  and
      Wang, Bixuan  and
      Zhang, Yifei",
    editor = "Che, Wanxiang  and
      Nabende, Joyce  and
      Shutova, Ekaterina  and
      Pilehvar, Mohammad Taher",
    booktitle = "Findings of the Association for Computational Linguistics: ACL 2025",
    month = jul,
    year = "2025",
    address = "Vienna, Austria",
    publisher = "Association for Computational Linguistics",
    url = "https://aclanthology.org/2025.findings-acl.1192/",
    doi = "10.18653/v1/2025.findings-acl.1192",
    pages = "23221--23235",
    ISBN = "979-8-89176-256-5"
}

@inproceedings{gaur-etal-2025-assess,
    title = "Assess and Prompt: A Generative {RL} Framework for Improving Engagement in Online Mental Health Communities",
    author = "Gaur, Bhagesh  and
      Gupta, Karan  and
      Srivastava, Aseem  and
      Gupta, Manish  and
      Akhtar, Md Shad",
    editor = "Christodoulopoulos, Christos  and
      Chakraborty, Tanmoy  and
      Rose, Carolyn  and
      Peng, Violet",
    booktitle = "Findings of the Association for Computational Linguistics: EMNLP 2025",
    month = nov,
    year = "2025",
    address = "Suzhou, China",
    publisher = "Association for Computational Linguistics",
    url = "https://aclanthology.org/2025.findings-emnlp.982/",
    doi = "10.18653/v1/2025.findings-emnlp.982",
    pages = "18102--18118",
    ISBN = "979-8-89176-335-7"
}

@inproceedings{nguyen-etal-2025-hanging,
    title = "Hanging in the Balance: Pivotal Moments in Crisis Counseling Conversations",
    author = "Nguyen, Vivian  and
      Lee, Lillian  and
      Danescu-Niculescu-Mizil, Cristian",
    editor = "Che, Wanxiang  and
      Nabende, Joyce  and
      Shutova, Ekaterina  and
      Pilehvar, Mohammad Taher",
    booktitle = "Proceedings of the 63rd Annual Meeting of the Association for Computational Linguistics (Volume 1: Long Papers)",
    month = jul,
    year = "2025",
    address = "Vienna, Austria",
    publisher = "Association for Computational Linguistics",
    url = "https://aclanthology.org/2025.acl-long.1440/",
    doi = "10.18653/v1/2025.acl-long.1440",
    pages = "29801--29817",
    ISBN = "979-8-89176-251-0"
}

@inproceedings{qiu-etal-2025-deepwell,
    title = "{D}eep{W}ell-Adol: A Scalable Expert-Based Dialogue Corpus for Adolescent Positive Mental Health and Wellbeing Promotion",
    author = "Qiu, Wenyu  and
      Wang, Yuxiong  and
      Tan, Jiajun  and
      Hou, Hanchao  and
      Liu, Qinda  and
      Yao, Wei  and
      Ni, Shiguang",
    editor = "Christodoulopoulos, Christos  and
      Chakraborty, Tanmoy  and
      Rose, Carolyn  and
      Peng, Violet",
    booktitle = "Proceedings of the 2025 Conference on Empirical Methods in Natural Language Processing",
    month = nov,
    year = "2025",
    address = "Suzhou, China",
    publisher = "Association for Computational Linguistics",
    url = "https://aclanthology.org/2025.emnlp-main.646/",
    doi = "10.18653/v1/2025.emnlp-main.646",
    pages = "12797--12821",
    ISBN = "979-8-89176-332-6"
}

@inproceedings{mahmood-etal-2025-fully,
    title = "A Fully Generative Motivational Interviewing Counsellor Chatbot for Moving Smokers Towards the Decision to Quit",
    author = "Mahmood, Zafarullah  and
      Ali, Soliman  and
      Zhu, Jiading  and
      Abdelwahab, Mohamed  and
      Collins, Michelle Yu  and
      Chen, Sihan  and
      Zhao, Yi Cheng  and
      Wolff, Jodi  and
      Melamed, Osnat C.  and
      Minian, Nadia  and
      Maslej, Marta  and
      Cooper, Carolynne  and
      Ratto, Matt  and
      Selby, Peter  and
      Rose, Jonathan",
    editor = "Che, Wanxiang  and
      Nabende, Joyce  and
      Shutova, Ekaterina  and
      Pilehvar, Mohammad Taher",
    booktitle = "Findings of the Association for Computational Linguistics: ACL 2025",
    month = jul,
    year = "2025",
    address = "Vienna, Austria",
    publisher = "Association for Computational Linguistics",
    url = "https://aclanthology.org/2025.findings-acl.1283/",
    doi = "10.18653/v1/2025.findings-acl.1283",
    pages = "25008--25043",
    ISBN = "979-8-89176-256-5"
}

@inproceedings{lee-etal-2024-detecting-bipolar,
    title = "Detecting Bipolar Disorder from Misdiagnosed Major Depressive Disorder with Mood-Aware Multi-Task Learning",
    author = "Lee, Daeun  and
      Jeon, Hyolim  and
      Son, Sejung  and
      Park, Chaewon  and
      An, Ji hyun  and
      Kim, Seungbae  and
      Han, Jinyoung",
    editor = "Duh, Kevin  and
      Gomez, Helena  and
      Bethard, Steven",
    booktitle = "Proceedings of the 2024 Conference of the North American Chapter of the Association for Computational Linguistics: Human Language Technologies (Volume 1: Long Papers)",
    month = jun,
    year = "2024",
    address = "Mexico City, Mexico",
    publisher = "Association for Computational Linguistics",
    url = "https://aclanthology.org/2024.naacl-long.278/",
    doi = "10.18653/v1/2024.naacl-long.278",
    pages = "4954--4970"
}

@misc{bar-shahar2025,
  author       = {Bar-Shachar, Yael and Rafael, Dana and Goel, Anmol and Klein, Ayal and Gurevych, Iryna and Atzil-Slonim, Dana},
  title        = {Evaluating the Capabilities of Large Language Models (LLMs) versus Human Therapists to Generate Personalized Interventions},
  year         = {2025},
  note         = {Preprint on the Open Science Framework},
url={https://osf.io/pjhak/overview?view_only=2b72405451d54c33a4776aa2b20cb0c4}
}

@inproceedings{singla-etal-2020-towards,
    title = "Towards end-2-end learning for predicting behavior codes from spoken utterances in psychotherapy conversations",
    author = "Singla, Karan  and
      Chen, Zhuohao  and
      Atkins, David  and
      Narayanan, Shrikanth",
    editor = "Jurafsky, Dan  and
      Chai, Joyce  and
      Schluter, Natalie  and
      Tetreault, Joel",
    booktitle = "Proceedings of the 58th Annual Meeting of the Association for Computational Linguistics",
    month = jul,
    year = "2020",
    address = "Online",
    publisher = "Association for Computational Linguistics",
    url = "https://aclanthology.org/2020.acl-main.351/",
    doi = "10.18653/v1/2020.acl-main.351",
    pages = "3797--3803"
}

@inproceedings{cheng-etal-2023-pal,
    title = "{PAL}: Persona-Augmented Emotional Support Conversation Generation",
    author = "Cheng, Jiale  and
      Sabour, Sahand  and
      Sun, Hao  and
      Chen, Zhuang  and
      Huang, Minlie",
    editor = "Rogers, Anna  and
      Boyd-Graber, Jordan  and
      Okazaki, Naoaki",
    booktitle = "Findings of the Association for Computational Linguistics: ACL 2023",
    month = jul,
    year = "2023",
    address = "Toronto, Canada",
    publisher = "Association for Computational Linguistics",
    url = "https://aclanthology.org/2023.findings-acl.34/",
    doi = "10.18653/v1/2023.findings-acl.34",
    pages = "535--554"
}

@article{song2026mentalbench,
  title={MentalBench: A Benchmark for Evaluating Psychiatric Diagnostic Capability of Large Language Models},
  author={Song, Hoyun and Kang, Migyeong and Shin, Jisu and Kim, Jihyun and Park, Chanbi and Yoo, Hangyeol and An, Jihyun and Oh, Alice and Han, Jinyoung and Lim, KyungTae},
  journal={arXiv preprint arXiv:2602.12871},
  year={2026}
}

\appendix

\section{Details about the surveyed papers}
\label{sec:acldetails}

The full list of surveyed papers with the observed practices are in Tables~\ref{tab:surveyed_works}--\ref{tab:surveyed_works_14}. Psychologically grounded metrics are based on criteria derived from psychological theory, clinical research, or expert input. AI/NLP metrics, by contrast, focus on computational performance (e.g., accuracy, BLEU, ROUGE) and are largely agnostic to psychological or therapeutic soundness. 

The tasks addressed in these papers are listed in Table~\ref{tab:mh_tasks}. 

Some papers examine mental disorders at a broad level, while others focus on specific diagnosed conditions. Among those that target specific conditions, the following named disorders are examined: \textit{Anxiety, Depression, Suicide ideation, Cognitive distortions, Post-Traumatic Stress Disorder (PTSD), Bipolar Disorder, Schizophrenia, Self-harm, Anorexia, Trauma, Stress, Attention-Deficit/Hyperactivity Disorder (ADHD), Obsessive-Compulsive Disorder (OCD), Panic, and Addiction}.

\begin{table*}[t]
\centering
\begin{tabular}{p{5cm}p{2cm}p{1.7cm}p{1.7cm}p{1.7cm}p{1.7cm}}
\hline
Paper  & Evaluation metrics  &  Human evaluation & Expert evaluators & Evaluation guidelines provided & Discuss limitations of evaluation \\
\hline
Social Biases in NLP Models as Barriers for Persons with Disabilities~\cite{hutchinson-etal-2020-social} &  AI/NLP metrics & No & No &  N.\,A. & Yes\\
Suicidal Risk Detection for Military Personnel~\cite{park2020suicidal} & AI/NLP metrics &  No & No & N.\,A. & No\\
Towards end-2-end learning for predicting behavior codes from spoken utterances in psychotherapy conversations~\cite{singla-etal-2020-towards} &  AI/NLP metrics & No & No &  N.\,A. & No\\
Cross-Lingual Suicidal-Oriented Word Embedding toward Suicide Prevention~\cite{lee-etal-2020-cross} & AI/NLP metrics & No & No & N.\,A. & No\\
A Computational Approach to Understanding Empathy Expressed in Text-Based Mental Health Support~\cite{sharma-etal-2020-computational} & AI/NLP metrics & No & No & N.\,A. & No\\
Do Models of Mental Health Based on Social Media Data Generalize?~\cite{harrigian-etal-2020-models} & AI/NLP metrics & No & No & N.\,A. & No\\
Suicide Ideation Detection via Social and Temporal User Representations using Hyperbolic Learning~\cite{sawhney-etal-2021-suicide} & AI/NLP metrics & No & No & N.\,A. & Yes\\
Empirical Evaluation of Pre-trained Transformers for Human-Level NLP: The Role of Sample Size and Dimensionality~\cite{v-ganesan-etal-2021-empirical} & AI/NLP metrics & No & No &  N.\,A. & No\\
Gender and Racial Fairness in Depression Research using Social Media~\cite{aguirre-etal-2021-gender} & AI/NLP metrics & No & No &  N.\,A. & Yes\\
Development of Conversational AI for Sleep Coaching Programme~\cite{shim-2021-development} & AI/NLP metrics & No & No & N.\,A. & No\\
\hline
\end{tabular}
\caption{List of surveyed papers \textit{(part 1 of 14)}.}
\label{tab:surveyed_works}
\end{table*}

\begin{table*}[t]
\centering
\begin{tabular}{p{5cm}p{2.2cm}p{1.7cm}p{1.7cm}p{1.7cm}p{1.7cm}}
\hline
Paper & Evaluation metrics  &  Human evaluation & Expert evaluators & Evaluation guidelines provided & Discuss limitations of evaluation \\
\hline
Linguistic Complexity Loss in Text-Based Therapy~\cite{wei2021linguistic} & AI/NLP metrics  &No &No & N.\,A. & Yes\\
Weakly-Supervised Methods for Suicide Risk Assessment: Role of Related Domains~\cite{yang-etal-2021-weakly} & AI/NLP metrics & No & No & N.\,A. & Yes \\
Towards Emotional Support Dialog Systems~\cite{liu2021towards} & Psychologically grounded metrics & Yes & No & Yes & Yes\\
PHASE: Learning Emotional Phase-aware Representations for Suicide Ideation Detection on Social Media~\cite{sawhney-etal-2021-phase} & AI/NLP metrics & No & No & N.\,A.  & Yes\\
Micromodels for Efficient, Explainable, and Reusable Systems: A Case Study on Mental Health~\cite{lee-etal-2021-micromodels-efficient} & AI/NLP metrics & No & No  &  N.\,A. & No\\
Predicting Treatment Outcome from Patient Texts:The Case of Internet-Based Cognitive Behavioural Therapy~\cite{gogoulou-etal-2021-predicting} & AI/NLP metrics & No & No &  N.\,A. & Yes\\
Towards Intelligent Clinically-Informed Language Analyses of People with Bipolar Disorder and Schizophrenia~\cite{aich-etal-2022-towards} & AI/NLP metrics& No & No & N.\,A. & Yes\\
Identifying Moments of Change from Longitudinal User Text~\cite{tsakalidis-etal-2022-identifying} &  AI/NLP metrics & Yes & No &  Yes & No\\
Improving the Generalizability of Depression Detection by Leveraging Clinical Questionnaires~\cite{nguyen-etal-2022-improving} & AI/NLP metrics & No & No & N.\,A. & Yes\\
A Shoulder to Cry on: Towards A Motivational Virtual Assistant for Assuaging Mental Agony~\cite{saha-etal-2022-shoulder} & AI/NLP metrics & Yes & No & No & No\\
\hline
\end{tabular}
\caption{List of surveyed papers \textit{(part 2 of 14)}.}
\label{tab:surveyed_works_2}
\end{table*}

\begin{table*}[t]
\centering
\begin{tabular}{p{5cm}p{2.2cm}p{1.7cm}p{1.7cm}p{1.7cm}p{1.7cm}}
\hline
Paper & Evaluation metrics  &  Human evaluation & Expert evaluators & Evaluation guidelines provided & Discuss limitations of evaluation \\
\hline
D4: a Chinese Dialogue Dataset for Depression-Diagnosis-Oriented Chat~\cite{yao-etal-2022-d4} & Psychologically grounded metrics & Yes &  Yes & No & Yes\\
Gendered Mental Health Stigma in Masked Language Models~\cite{lin-etal-2022-gendered} & AI/NLP metrics & No & No &  N.\,A. & Yes\\
Leveraging Open Data and Task Augmentation to Automated Behavioral Coding of Psychotherapy Conversations in Low-Resource Scenarios~\cite{chen-etal-2022-leveraging-open} & AI/NLP metrics & No & No &  N.\,A. & No\\
PAIR: Prompt-Aware margIn Ranking for Counselor Reflection Scoring in Motivational Interviewing~\cite{min-etal-2022-pair} & Psychologically grounded metrics & Yes & Yes  & Yes & Yes\\
Symptom Identification for Interpretable Detection of Multiple Mental Disorders on Social Media~\cite{zhang-etal-2022-symptom} & AI/NLP metrics & No & No &  N.\,A. & Yes\\
An Annotated Dataset for Explainable Interpersonal Risk Factors of Mental Disturbance in Social Media Posts~\cite{garg-etal-2023-annotated} & AI/NLP metrics& No & No & N.\,A. &  No\\
C2D2 Dataset: A Resource for the Cognitive Distortion Analysis and Its Impact on Mental Health~\cite{wang-etal-2023-c2d2} & AI/NLP metrics & No & No & N.\,A. & Yes\\
PAL: Persona-Augmented Emotional Support Conversation Generation~\cite{cheng-etal-2023-pal} & Psychologically grounded metrics & Yes & No & Yes & Yes\\
Discourse-Level Representations can Improve Prediction of Degree of Anxiety~\cite{juhng-etal-2023-discourse}& AI/NLP metrics & No & No &  N.\,A. & Yes\\
Ask an Expert: Leveraging Language Models to Improve Strategic Reasoning in Goal-Oriented Dialogue Models~\cite{zhang-etal-2023-ask} & Psychologically grounded metrics & Yes & No & Yes & Yes\\
\hline
\end{tabular}
\caption{List of surveyed papers \textit{(part 3 of 14)}.}
\label{tab:surveyed_works_3}
\end{table*}

\begin{table*}[t]
\centering
\begin{tabular}{p{5.2cm}p{2.2cm}p{1.7cm}p{1.7cm}p{1.7cm}p{1.7cm}}
\hline
Paper & Evaluation metrics  &  Human evaluation & Expert evaluators & Evaluation guidelines provided & Discuss limitations of evaluation \\
\hline
Empowering Psychotherapy with Large Language Models: Cognitive Distortion Detection through Diagnosis of Thought Prompting~\cite{chen-etal-2023-empowering} &  AI/NLP metrics & No & No & N.\,A. & Yes\\
Identifying Early Maladaptive Schemas from Mental Health Question Texts~\cite{gollapalli-etal-2023-identifying} & AI/NLP metrics & No & No &  N.\,A. & Yes\\
What to Fuse and How to Fuse: Exploring Emotion and Personality Fusion Strategies for Explainable Mental Disorder Detection~\cite{zanwar-etal-2023-fuse} & AI/NLP metrics & No & No & N.\,A. & No\\
Understanding Client Reactions in Online Mental Health Counseling~\cite{li-etal-2023-understanding} &  AI/NLP metrics & No & No & N.\,A. & Yes\\
Cognitive Reframing of Negative Thoughts through Human-Language Model Interaction~\cite{sharma-etal-2023-cognitive} & Psychologically grounded metrics & Yes & Yes & Yes & Yes\\
Knowledge-enhanced Mixed-initiative Dialogue System for Emotional Support Conversations~\cite{deng-etal-2023-knowledge} & Psychologically grounded metrics & Yes & No & No & Yes\\
Language and Mental Health: Measures of Emotion Dynamics from Text as Linguistic Biosocial Markers~\cite{teodorescu-etal-2023-language} & AI/NLP metrics & No & No &  N.\,A. & Yes\\
A Cognitive Stimulation Dialogue System with Multi-source Knowledge Fusion for Elders with Cognitive Impairment~\cite{jiang-etal-2023-cognitive} & Psychologically grounded metrics & Yes & No & No & No\\
A Simple and Flexible Modeling for Mental Disorder Detection by Learning from Clinical Questionnaires~\cite{song-etal-2023-simple} & AI/NLP metrics & No & No &   N.\,A. & Yes\\
Task-Adaptive Tokenization: Enhancing Long-Form Text Generation Efficacy in Mental Health and Beyond~\cite{liu-etal-2023-task} & Psychologically grounded metrics & Yes & Yes & Yes & No\\
\hline
\end{tabular}
\caption{List of surveyed papers \textit{(part 4 of 14)}.}
\label{tab:surveyed_works_4}
\end{table*}

\begin{table*}[t]
\centering
\begin{tabular}{p{5cm}p{2.2cm}p{1.7cm}p{1.7cm}p{1.7cm}p{1.7cm}}
\hline
Paper & Evaluation metrics  &  Human evaluation & Expert evaluators & Evaluation guidelines provided & Discuss limitations of evaluation \\
\hline
Self-Adapted Utterance Selection for Suicidal Ideation Detection in Lifeline Conversations~\cite{wang-etal-2023-self} & AI/NLP metrics & No & No &  N.\,A. & No\\
FedTherapist: Mental Health Monitoring with User-Generated Linguistic Expressions on Smartphones via Federated Learning~\cite{shin-etal-2023-fedtherapist} & AI/NLP metrics & No & No & N.\,A. & Yes\\
Towards Interpretable Mental Health Analysis with Large Language Models~\cite{yang-etal-2023-towards} & AI/NLP metrics  & Yes & No & Yes & No\\
SMHD-GER: A Large-Scale Benchmark Dataset for Automatic Mental Health Detection from Social Media in German~\cite{zanwar-etal-2023-smhd} & AI/NLP metrics  & No & No &  N.\,A. & No\\
Towards Identifying Fine-Grained Depression Symptoms from Memes~\cite{yadav-etal-2023-towards} & AI/NLP metrics & Yes & No & No & Yes\\
e-THERAPIST: I suggest you to cultivate a mindset of positivity and nurture uplifting thoughts~\cite{mishra-etal-2023-e} & Psychologically grounded metrics & Yes & No & No & No\\
Sequential Path Signature Networks for Personalised Longitudinal Language Modeling~\cite{tseriotou-etal-2023-sequential} & AI/NLP metrics & Yes & No & Yes & No\\
DisorBERT: A Double Domain Adaptation Model for Detecting Signs of Mental Disorders in Social Media~\cite{aragon-etal-2023-disorbert} & AI/NLP metrics & No & No  & N.\,A. & No\\
PAL to Lend a Helping Hand: Towards Building an Emotion Adaptive Polite and Empathetic Counseling Conversational Agent~\cite{mishra-etal-2023-pal} & Psychologically grounded metrics & Yes & Yes & No & No\\
Detection of Multiple Mental Disorders from Social Media with Two-Stream Psychiatric Experts~\cite{chen-etal-2023-detection} & AI/NLP metrics & No & No & N.\,A. & No\\
\hline
\end{tabular}
\caption{List of surveyed papers \textit{(part 5 of 14)}.}
\label{tab:surveyed_works_5}
\end{table*}

\begin{table*}[t]
\centering
\begin{tabular}{p{5.1cm}p{2.2cm}p{1.7cm}p{1.7cm}p{1.7cm}p{1.7cm}}
\hline
Paper & Evaluation metrics  &  Human evaluation & Expert evaluators & Evaluation guidelines provided & Discuss limitations of evaluation \\
\hline
Depression Detection in Clinical Interviews with LLM-Empowered Structural Element Graph~\cite{chen-etal-2024-depression} & AI/NLP metrics  & No & No & N.\,A.& No\\
Generating Mental Health Transcripts with SAPE (Spanish Adaptive Prompt Engineering)~\cite{lozoya-etal-2024-generating} & Psychologically grounded metrics & Yes & Yes & No &  Yes\\
Taking a turn for the better: Conversation redirection throughout the course of mental-health therapy~\cite{nguyen-etal-2024-taking} & Psychologically grounded metrics & Yes & No & No & Yes\\
Detecting Bipolar Disorder from Misdiagnosed Major Depressive Disorder with Mood-Aware Multi-Task Learning~\cite{lee-etal-2024-detecting-bipolar} & AI/NLP metrics & No & No & N.\,A. & Yes\\
Mental Disorder Classification via Temporal Representation of Text~\cite{kumar-etal-2024-mental} & AI/NLP metrics & No & No & N.\,A. & No\\
Multi-Level Feedback Generation with LLMs for Empowering Novice Peer Counselors~\cite{chaszczewicz-etal-2024-multi} & Psychologically grounded metrics & Yes & Yes & Yes & Yes\\
Understanding the Therapeutic Relationship between Counselors and Clients in Online Text-based Counseling using LLMs~\cite{li-etal-2024-understanding-therapeutic} & Psychologically grounded metrics & Yes & Yes & Yes & Yes\\
Diverse Perspectives, Divergent Models: Cross-Cultural Evaluation of Depression Detection on Twitter~\cite{abdelkadir-etal-2024-diverse} & AI/NLP metrics & No & No & N.\,A. & No\\
IMBUE: Improving Interpersonal Effectiveness through Simulation and Just-in-time Feedback with Human-Language Model Interaction~\cite{lin-etal-2024-imbue} & Psychologically grounded metrics & Yes & Yes & Yes & Yes\\
PsychoGAT: A Novel Psychological Measurement Paradigm through Interactive Fiction Games with LLM Agents~\cite{yang-etal-2024-psychogat} & Psychologically grounded metrics & Yes & Yes & No & Yes\\
\hline
\end{tabular}
\caption{List of surveyed papers \textit{(part 6 of 14)}.}
\label{tab:surveyed_works_6}
\end{table*}

\begin{table*}[t]
\centering
\begin{tabular}{p{5cm}p{2.2cm}p{1.7cm}p{1.7cm}p{1.7cm}p{1.7cm}}
\hline
Paper & Evaluation metrics  &  Human evaluation & Expert evaluators & Evaluation guidelines provided & Discuss limitations of evaluation \\
\hline
ALBA: Adaptive Language-Based Assessments for Mental Health~\cite{varadarajan-etal-2024-alba} & AI/NLP metrics& No & No & N.\,A.  & Yes\\
Ask the experts: sourcing a high-quality nutrition counseling dataset through Human-AI collaboration~\cite{balloccu-etal-2024-ask} & Psychologically grounded metrics & Yes & Yes  & Yes & Yes\\ 
CURE: Context- and Uncertainty-Aware Mental Disorder Detection~\cite{kang-etal-2024-cure} & AI/NLP metrics & No & No & N.\,A. & Yes\\
Still Not Quite There! Evaluating Large Language Models for Comorbid Mental Health Diagnosis~\cite{hengle-etal-2024-still} & AI/NLP metrics & No & No &   N.\,A. & Yes\\
Combining Hierachical VAEs with LLMs for clinically meaningful timeline summarisation in social media~\cite{song-etal-2024-combining} & Psychologically grounded metrics & Yes & Yes & Yes & Yes\\
On the Way to Gentle AI Counselor: Politeness Cause Elicitation and Intensity Tagging in Code-mixed Hinglish Conversations for Social Good~\cite{priya-etal-2024-way} & AI/NLP metrics & No & No &  N.\,A. & Yes\\
LLM Questionnaire Completion for Automatic Psychiatric Assessment~\cite{rosenman-etal-2024-llm} & AI/NLP metrics & No & No & N.\,A. & Yes\\
Emotion Granularity from Text: An Aggregate-Level Indicator of Mental Health~\cite{vishnubhotla-etal-2024-emotion-granularity} & AI/NLP metrics & No & No & N.\,A. & Yes\\
CASE: Efficient Curricular Data Pre-training for Building Assistive Psychology Expert Models~\cite{harne-etal-2024-case} & AI/NLP metrics & No & No &  N.\,A. & No\\
Can AI Relate: Testing Large Language Model Response for Mental Health Support~\cite{gabriel-etal-2024-ai} & Psychologically grounded metrics & Yes & Yes & Yes & Yes\\
\hline
\end{tabular}
\caption{List of surveyed papers \textit{(part 7 of 14)}.}
\label{tab:surveyed_works_7}
\end{table*}

\begin{table*}[t]
\centering
\begin{tabular}{p{5cm}p{2.2cm}p{1.7cm}p{1.7cm}p{1.7cm}p{1.7cm}}
\hline
Paper & Evaluation metrics  &  Human evaluation & Expert evaluators & Evaluation guidelines provided & Discuss limitations of evaluation \\
\hline
Crisis counselor language and perceived genuine concern in crisis conversations~\cite{buda-etal-2024-crisis} & AI/NLP metrics & No & No & N.\,A. & Yes\\
Roleplay-doh: Enabling Domain-Experts to Create LLM-simulated Patients via Eliciting and Adhering to Principles~\cite{louie-etal-2024-roleplay} & Psychologically grounded metrics & Yes & Yes & Yes & Yes\\
Exciting Mood Changes: A Time-aware Hierarchical Transformer for Change Detection Modelling~\cite{hills-etal-2024-exciting} & AI/NLP metrics & Yes & No &  Yes  & No\\
SMILE: Single-turn to Multi-turn Inclusive Language Expansion via ChatGPT for Mental Health Support~\cite{qiu-etal-2024-smile} & Psychologically grounded metrics & Yes & Yes & Yes & Yes\\
PsyGUARD: An Automated System for Suicide Detection and Risk Assessment in Psychological Counseling~\cite{qiu-etal-2024-psyguard} & AI/NLP metrics & No & No & N.\,A.  & No\\
Modeling Empathetic Alignment in Conversation~\cite{yang-jurgens-2024-modeling} & AI/NLP metrics & No & No &  N.\,A. & Yes\\
Mapping Long-term Causalities in Psychiatric Symptomatology and Life Events from Social Media~\cite{chen-etal-2024-mapping} & AI/NLP metrics & No & No &   N.\,A. & No\\
Knowledge Planning in LLMs for Domain-Aligned Counseling Summarization~\cite{srivastava-etal-2024-knowledge} & Psychologically grounded metrics & Yes & Yes & Yes & Yes\\
Using LLMs to Simulate Patients for Training Mental Health Professionals~\cite{wang-etal-2024-patient}  &  Psychologically grounded metrics & Yes & Yes & Yes & Yes\\
Deciphering Cognitive Distortions in Patient-Doctor Mental Health Conversations: A Multimodal LLM-Based Detection and Reasoning Framework~\cite{singh-etal-2024-deciphering} & AI/NLP metrics & Yes & No & Yes & Yes\\
\hline
\end{tabular}
\caption{List of surveyed papers \textit{(part 8 of 14)}.}
\label{tab:surveyed_works_8}
\end{table*}

\begin{table*}[t]
\centering
\begin{tabular}{p{5cm}p{2.2cm}p{1.7cm}p{1.7cm}p{1.7cm}p{1.7cm}}
\hline
Paper & Evaluation metrics  &  Human evaluation & Expert evaluators & Evaluation guidelines provided & Discuss limitations of evaluation \\
\hline
Cactus: Towards Psychological Counseling Conversations using Cognitive Behavioral Theory~\cite{lee-etal-2024-cactus} & Psychologically grounded metrics & Yes & Yes & Yes & No\\
Chinese MentalBERT: Domain-Adaptive Pre-training on Social Media for Chinese Mental Health Text Analysis~\cite{zhai-etal-2024-chinese} & AI/NLP metrics & No & No &  N.\,A. & No\\
HealMe: Harnessing Cognitive Reframing in LLMs for Psychotherapy~\cite{xiao-etal-2024-healme} & Psychologically grounded metrics & Yes & Yes & Yes & No\\
When LLMs Meets Acoustic Landmarks: An Efficient Approach to Integrate Speech into Large Language Models for Depression Detection~\cite{zhang-etal-2024-llms} & AI/NLP metrics & No & No & N.\,A. & No \\
The Colorful Future of LLMs: Evaluating and Improving LLMs as Emotional Supporters for Queer Youth~\cite{lissak-etal-2024-colorful} & Psychologically grounded metrics & Yes & No & Yes & No\\
Decoding the Narratives: Analyzing Personal Drug Experiences Shared on Reddit~\cite{bouzoubaa-etal-2024-decoding} & AI/NLP metrics & No & No & N.\,A. & Yes\\
Lived Experience Not Found: LLMs Struggle to Align with Experts on Addressing Adverse Drug Reactions from Psychiatric Medication Use~\cite{chandra-etal-2025-lived} & Psychologically grounded metrics & Yes & Yes & Yes & Yes\\
Using Linguistic Entrainment to Evaluate LLMs for Use in Cognitive Behavioral Therapy~\cite{kian-etal-2025-using} & AI/NLP metrics & No & No &   N.\,A. & Yes\\
Do Large Language Models Align with Core Mental Health Counseling Competencies?~\cite{nguyen-etal-2025-large} & AI/NLP metrics & No & No &  N.\,A. & Yes\\
KMI: A Dataset of Korean Motivational Interviewing Dialogues for Psychotherapy~\cite{kim-etal-2025-kmi} & Psychologically grounded metrics & Yes & Yes & Yes & No\\
\hline
\end{tabular}
\caption{List of surveyed papers \textit{(part 9 of 14)}.}
\label{tab:surveyed_works_9}
\end{table*}

\begin{table*}[t]
\centering
\begin{tabular}{p{5.3cm}p{2.2cm}p{1.7cm}p{1.7cm}p{1.7cm}p{1.7cm}}
\hline
Paper & Evaluation metrics  &  Human evaluation & Expert evaluators & Evaluation guidelines provided & Discuss limitations of evaluation \\
\hline
CBT-Bench: Evaluating Large Language Models on Assisting Cognitive Behavior Therapy~\cite{zhang-etal-2025-cbt} & Psychologically grounded metrics & Yes & Yes & Yes & Yes\\
Multimodal Cognitive Reframing Therapy via Multi-hop Psychotherapeutic Reasoning~\cite{kim-etal-2025-multimodal} &  Psychologically grounded metrics & Yes & Yes & Yes & Yes\\
A Fully Generative Motivational Interviewing Counsellor Chatbot for Moving Smokers Towards the Decision to Quit~\cite{mahmood-etal-2025-fully} & Psychologically grounded metrics  &  Yes &  No & Yes & Yes\\
PsyDial: A Large-scale Long-term Conversational Dataset for Mental Health Support~\cite{qiu-lan-2025-psydial}  & Psychologically grounded metrics  & Yes & Yes & Yes & Yes \\
SpeechT-RAG: Reliable Depression Detection in LLMs with Retrieval-Augmented Generation Using Speech Timing Information~\cite{zhang-etal-2025-speecht}  & AI/NLP metrics & No  &No  & N.\,A.& No \\
DeepWell-Adol: A Scalable Expert-Based Dialogue Corpus for Adolescent Positive Mental Health and Wellbeing Promotion~\cite{qiu-etal-2025-deepwell}  & Psychologically grounded metrics  & Yes  &Yes  & Yes& Yes \\
Hanging in the Balance: Pivotal Moments in Crisis Counseling Conversations~\cite{nguyen-etal-2025-hanging}& Psychologically grounded metrics& No & No & N.\,A.& Yes\\
Assess and Prompt: A Generative RL Framework for Improving Engagement in Online Mental Health Communities~\cite{gaur-etal-2025-assess} & AI/NLP metrics  & Yes &  No &  No & Yes \\
AnnaAgent: Dynamic Evolution Agent System with Multi-Session Memory for Realistic Seeker Simulation~\cite{wang-etal-2025-annaagent}&  Psychologically grounded metrics& No & No  &N.\,A. & Yes\\
Tracking Life's Ups and Downs: Mining Life Events from Social Media Posts for Mental Health Analysis~\cite{lv-etal-2025-tracking} & Psychologically grounded metrics&  No& No &  N.\,A.&Yes \\
\hline
\end{tabular}
\caption{List of surveyed papers \textit{(part 10 of 14)}.}
\label{tab:surveyed_works_10}
\end{table*}

\begin{table*}[t]
\centering
\begin{tabular}{p{5.3cm}p{2.2cm}p{1.7cm}p{1.7cm}p{1.7cm}p{1.7cm}}
\hline
Paper & Evaluation metrics  &  Human evaluation & Expert evaluators & Evaluation guidelines provided & Discuss limitations of evaluation \\
\hline
Can Large Language Models Identify Implicit Suicidal Ideation? An Empirical Evaluation~\cite{li-etal-2025-large-language-models-identify} &Psychologically grounded metrics & Yes &  Yes & Yes& No \\
Eeyore: Realistic Depression Simulation via Expert-in-the-Loop Supervised and Preference Optimization~\cite{liu-etal-2025-eeyore} & Psychologically grounded metrics& Yes & Yes & Yes& Yes\\
Dialogue Systems for Emotional Support via Value Reinforcement~\cite{kim-etal-2025-dialogue}  &Psychologically grounded metrics &  Yes&  Yes& Yes& Yes\\
MultiAgentESC: A LLM-based Multi-Agent Collaboration Framework for Emotional Support Conversation~\cite{xu-etal-2025-multiagentesc}  & Psychologically grounded metrics&  Yes& Yes & Yes& No\\
MAGI: Multi-Agent Guided Interview for Psychiatric Assessment~\cite{bi-etal-2025-magi} &Psychologically grounded metrics & Yes & Yes &Yes & Yes\\
Systematic Evaluation of Auto-Encoding and Large Language Model Representations for Capturing Author States and Traits~\cite{singh-etal-2025-systematic}  & AI/NLP metrics& No & No & N.\,A.& No \\
The Pursuit of Empathy: Evaluating Small Language Models for PTSD Dialogue Support~\cite{bn-etal-2025-pursuit}   &Psychologically grounded metrics & Yes  &No  &Yes & Yes\\
Just a Scratch: Enhancing LLM Capabilities for Self-harm Detection through Intent Differentiation and Emoji Interpretation~\cite{ghosh-etal-2025-just} &AI/NLP metrics &  No&No  & N.\,A.& Yes \\
Temporal reasoning for timeline summarisation in social media~\cite{song-etal-2025-temporal} & Psychologically grounded metrics & Yes &Yes  &Yes &No \\
MIRROR: Multimodal Cognitive Reframing Therapy for Rolling with Resistance~\cite{kim-etal-2025-mirror} &  Psychologically grounded metrics & Yes  &Yes  & Yes & Yes\\
\hline
\end{tabular}
\caption{List of surveyed papers \textit{(part 11 of 14)}.}
\label{tab:surveyed_works_11}
\end{table*}

\begin{table*}[t]
\centering
\begin{tabular}{p{5.5cm}p{2.2cm}p{1.7cm}p{1.7cm}p{1.7cm}p{1.7cm}}
\hline
Paper & Evaluation metrics  &  Human evaluation & Expert evaluators & Evaluation guidelines provided & Discuss limitations of evaluation \\
\hline
Are LLMs effective psychological assessors? Leveraging adaptive RAG for interpretable mental health screening through psychometric practice~\cite{ravenda-etal-2025-llms} &AI/NLP metrics &  No& No & N.\,A. & No \\
ReDepress: A Cognitive Framework for Detecting Depression Relapse from Social Media~\cite{agarwal-etal-2025-redepress} &AI/NLP metrics & No & No & N.\,A. & No\\
MentalGLM Series: Explainable LLMs for Mental Health Analysis on Chinese Social Media~\cite{zhai-etal-2025-mentalglm} &Psychologically grounded metrics & Yes & Yes &Yes &Yes \\
Towards AI-Assisted Psychotherapy: Emotion-Guided Generative Interventions~\cite{haydarov-etal-2025-towards}  & Psychologically grounded metrics & Yes & Yes & Yes&Yes \\
Mitigating Interviewer Bias in Multimodal Depression Detection: An Approach with Adversarial Learning and Contextual Positional Encoding~\cite{zhang-poellabauer-2025-mitigating} &AI/NLP metrics & No  &No  &N.\,A. & Yes\\
Explainable Depression Detection in Clinical Interviews with Personalized Retrieval-Augmented Generation~\cite{zhang-etal-2025-explainable} & AI/NLP metrics & No  &No  &N.\,A. &No \\
From Heart to Words: Generating Empathetic Responses via Integrated Figurative Language and Semantic Context Signals~\cite{lee-etal-2025-heart} &Psychologically grounded metrics & Yes  &No  & Yes&Yes \\
Reframe Your Life Story: Interactive Narrative Therapist and Innovative Moment Assessment with Large Language Models~\cite{feng-etal-2025-reframe}  &Psychologically grounded metrics &Yes  &Yes  &Yes  &Yes \\
From Conversation to Automation: Leveraging LLMs for Problem-Solving Therapy Analysis~\cite{aghakhani-etal-2025-conversation}  & AI/NLP metrics&  No& No & N.\,A.&Yes \\
Feel the Difference? A Comparative Analysis of Emotional Arcs in Real and LLM-Generated CBT Sessions~\cite{wang-etal-2025-feel}&Psychologically grounded metrics & No &No  & N.\,A.& Yes\\
\hline
\end{tabular}
\caption{List of surveyed papers \textit{(part 12 of 14)}.}
\label{tab:surveyed_works_12}
\end{table*}

\begin{table*}[t]
\centering
\begin{tabular}{p{5.5cm}p{2.2cm}p{1.7cm}p{1.7cm}p{1.7cm}p{1.7cm}}
\hline
Paper & Evaluation metrics  &  Human evaluation & Expert evaluators & Evaluation guidelines provided & Discuss limitations of evaluation \\
\hline
CAMI: A Counselor Agent Supporting Motivational Interviewing through State Inference and Topic Exploration~\cite{yang-etal-2025-cami} & Psychologically grounded metrics &Yes  &Yes  & Yes& Yes\\
Consistent Client Simulation for Motivational Interviewing-based Counseling~\cite{yang-etal-2025-consistent}  & Psychologically grounded metrics& Yes & Yes & Yes& No\\
Does Rationale Quality Matter? Enhancing Mental Disorder Detection via Selective Reasoning Distillation~\cite{song-etal-2025-rationale} & Psychologically grounded metrics&  Yes& Yes & Yes&Yes \\
Crisp: Cognitive Restructuring of Negative Thoughts through Multi-turn Supportive Dialogues~\cite{zhou-etal-2025-crisp} & Psychologically grounded metrics & Yes &No  &Yes & Yes\\
ProMind-LLM: Proactive Mental Health Care via Causal Reasoning with Sensor Data~\cite{zheng-etal-2025-promind} &Psychologically grounded metrics & Yes & Yes &No &Yes \\
EmoAgent: Assessing and Safeguarding Human-AI Interaction for Mental Health Safety~\cite{qiu-etal-2025-emoagent} & Psychologically grounded metrics& No &No  & N.\,A. & Yes\\
MIND: Towards Immersive Psychological Healing with Multi-Agent Inner Dialogue~\cite{chen-etal-2025-mind}&Psychologically grounded metrics & Yes &Yes  & Yes&Yes \\
PsyDT: Using LLMs to Construct the Digital Twin of Psychological Counselor with Personalized Counseling Style for Psychological Counseling~\cite{xie-etal-2025-psydt} &Psychologically grounded metrics &Yes  &Yes  &Yes & Yes\\
Third-Person Appraisal Agent: Simulating Human Emotional Reasoning in Text with Large Language Models~\cite{hong-etal-2025-third} &Psychologically grounded metrics & Yes & No  & Yes&No \\
CATCH: A Novel Data Synthesis Framework for High Therapy Fidelity and Memory-Driven Planning Chain of Thought in AI Counseling~\cite{chen-etal-2025-catch} &Psychologically grounded metrics & Yes  & Yes & Yes&No \\ 
\hline
\end{tabular}
\caption{List of surveyed papers \textit{(part 13 of 14)}.}
\label{tab:surveyed_works_13}
\end{table*}

\begin{table*}[t]
\centering
\begin{tabular}{p{5.5cm}p{2.2cm}p{1.7cm}p{1.7cm}p{1.7cm}p{1.7cm}}
\hline
Paper & Evaluation metrics  &  Human evaluation & Expert evaluators & Evaluation guidelines provided & Discuss limitations of evaluation \\
\hline
Assessment and manipulation of latent constructs in pre-trained language models using psychometric scales~\cite{reuben-etal-2025-assessment}& Psychologically grounded metrics& No & No &N.\,A. &No \\
How Real Are Synthetic Therapy Conversations? Evaluating Fidelity in Prolonged Exposure Dialogues~\cite{bn-etal-2025-real} & Psychologically grounded metrics & Yes &Yes  &Yes & Yes\\
KoACD: The First Korean Adolescent Dataset for Cognitive Distortion Analysis via Role-Switching Multi-LLM Negotiation~\cite{kim-kim-2025-koacd} &Psychologically grounded metrics & Yes & Yes &Yes &Yes \\
Exploring Large Language Models for Detecting Mental Disorders~\cite{kuzmin-etal-2025-exploring} & AI/NLP metrics&  No& No & N.\,A.&Yes \\
M-Help: Using Social Media Data to Detect Mental Health Help-Seeking Signals~\cite{sathvik-etal-2025-help} &AI/NLP metrics & No & No &  N.\,A.& Yes\\
\hline
\end{tabular}
\caption{List of surveyed papers \textit{(part 14 of 14)}.}
\label{tab:surveyed_works_14}
\end{table*}

\begin{table*}[ht]
\centering
\renewcommand{\arraystretch}{1.2}
\begin{tabular}{|p{3cm}|p{11cm}|}
\hline
\textbf{Support Type} & \textbf{Tasks} \\
\hline

\textbf{Assessment} &
Anxiety detection; Depression detection; Classification of interpersonal risk factors; Adverse drug reactions detection; Suicide risk detection; Cognitive distortion detection; Detection of schizophrenia disorders; Detecting bipolar disorder; Mental disorder classification; Predicting degree of anxiety; Detection of moments of change; Maladaptive schema detection; Cross-cultural evaluation of depression detection; Psychological profile generation; Measuring emotion granularity from text to detect mental health conditions; Detecting mood changes in social media users over time; Automatic detection of mental health conditions from social media posts in German; Identifying depression symptoms from memes; Multimodal LLM-based cognitive distortions detection; Personalized mood change detection from users' online text over time; Predicting treatment outcome in internet-based therapy; Classification of Reddit drug-use narratives into psychologically and socially meaningful categories; Chinese language model for psychological text analysis on social media; Evaluating how well depression detection models generalize across social media platforms; Identifying social biases toward disability in NLP models; Analyze fairness and bias in depression detection models on social media across gender and racial groups\\
\hline

\textbf{Intervention} &
Emotional support conversation generation; Using entrainment in CBT; Nutrition counseling; Synthetic dialogue generation for elders with cognitive impairment; Mental illness conditioned motivational dialogue generation; Generating motivational interviewing dialogues; Cognitive reframing; AI-assisted multimodal therapy; Evaluating how well large language models can assist cognitive behavioral therapy; Developing dialogue system for mental health support; Structured, empathetic cognitive reframing in psychotherapy; LLMs as emotional supporters for queer youth; Generating synthetic therapy transcripts; Enhancing long-form text generation for psychological question-answering; Evaluating whether LLMs can provide ethical, empathetic, and theory-grounded responses for mental health support\\
\hline

\textbf{Information synthesis} &
 Analysis of quality of therapy conversations; Behavior code prediction; Understanding the therapeutic relationship between counselors and clients; Evaluating LLM alignment with counseling competencies; Enhancing interpersonal skills; Analysis of client reactions in online mental health counseling; Understanding empathy in mental health support text; Clinically meaningful timeline summarisation in social media; Politeness and intensity tagging in conversations;   Teaching AI to automatically label behaviors in therapy conversations using small amounts of data; Scoring counselor responses for reflective listening in motivational interviewing; Creating realistic AI-simulated patients for counselor training; Counseling summarization; Patient simulation for training therapists   \\
\hline

\end{tabular}
\caption{Overview of the diverse tasks addressed in the surveyed papers.}
\label{tab:mh_tasks}
\end{table*}

\end{document}